\documentclass[12pt]{article}
\pdfoutput=1
\usepackage{amsfonts}
\usepackage{amssymb}
\usepackage{url}
\usepackage{hyperref}
\usepackage{simpler-wick}
\usepackage{bbold}
\usepackage{slashed}
\usepackage{graphicx}
\usepackage{pgfplots}
\usepackage{color}
\usepackage{mathrsfs}
\usepackage{fancybox}
\usepackage{lipsum}
\usepackage{eurosym}
\usepackage{tcolorbox}
\usepackage{lmodern}
\usepackage{tikz}
\usepackage{tikz,pgfplots}
\usepackage{pgfplots}
\usepackage{amsmath,amsthm,amssymb}
\pgfplotsset{compat=1.10}
\usepgfplotslibrary{fillbetween}
\usepackage{feynmp}
\usepackage{slashed}
\usepackage{graphicx}
\usepackage{epstopdf}
\usepackage{subfigure}
\usepackage{color}
\usetikzlibrary{decorations.pathmorphing}
\tikzset{snake it/.style={decorate, decoration=snake}}
\usetikzlibrary{shapes}
\usepackage{empheq}
\usepackage{lmodern}
\usepackage{anyfontsize}
\usepackage{t1enc}
\def\({\left (}
\def\){\right )}
\def\[{\left [}
\def\]{\right ]}
\def\d{\mathrm{d}}

\numberwithin{equation}{section}

\usepackage{environ}
\NewEnviron{myequation}{%
\begin{eqnarray}
\scalebox{1.05}{$\BODY$}
\end{eqnarray}
}

\newcommand{\ba}{\begin{equation}}
\newcommand{\ea}{\end{equation}}
 
\newcommand{\bea}{\begin{eqnarray}}

\newcommand{\bbbr}{\!\begin{array}}
\newcommand{\earr}{\end{array}\!}
\newcommand{\lb}{{\langle}}
\newcommand{\rb}{{\rangle}}

\newcommand{\be}{\begin{equation}}
\newcommand{\ee}{\end{equation}}

\def\lsim{\mathrel{\rlap{\lower4pt\hbox{\hskip1pt$\sim$}}
     \raise1pt\hbox{$<$}}}         
\def\gsim{\mathrel{\rlap{\lower4pt\hbox{\hskip1pt$\sim$}}
     \raise1pt\hbox{$>$}}}         

\begin{document}
\newcommand{\CC}{{\mbox{\textbf {\textit C}}}}
\newcommand{\RR}{{\mbox{\textbf {\textit R}}}}
\newcommand{\PP}{{\mbox{\textbf {\textit P}}}}
\newcommand{\nnn}{n}
\def\smallTFD{{\rm TFD}}
\def\d{{\partial}}
\def\n{{\bf \widehat n}}
\def\k{{\bf k}}
\def\changemargin#1#2{\list{}{\rightmargin#2\leftmargin#1}\item[]}
\let\endchangemargin=\endlist 
\def\smalll{\footnotesize}
\def\aaA{{{}_{\! \mbox{\scriptsize  $A$}}}}

\def\cC{\mbox{\textit{\textbf{C}\!\,}}}
\def\bBB{{{}_{\! \mbox{\scriptsize  $B$}}}}
\def\ccC{{{}_{\nspc \mbox{\scriptsize  $C$}}}}
\def\abAB{{{}_{\! \mbox{\scriptsize  $A\nspc\nspc B$}}}}
\def\bB{{\mbox{\scriptsize $b$}}}
\def\bbB{{\mbox{\scriptsize  $b$}}}

\def\ccc{{{\!}_{\mbox{\small {$c$}}}}}

\def\mn{{{}_{\raisebox{-.5pt}{\scriptsize $\mom \nom$}}}}
\def\nm{{{}_{\raisebox{-.5pt}{\scriptsize $\nom \mom$}}}}

\def\sS{\uU}

\def\bigll{\bigl}
\def\bigrr{\bigr}
\def\mathbi#1{\textbf{\em #1}} 
\def\som{{ \textit{\textbf s}}} 
\def\tom{{ \textit{\textbf t}}} 
\def\nom{{ \textit{\textbf n}}} 
\def\mom{{ \textit{\textbf m}}} 
\def\kom{{ \textit{\textbf k}}}  
\def\nomt{{ \textit{\textbf n}}}  
\def\momt{{ \textit{\textbf m}}}  
\def\komt{{ \textit{\textbf k}}} 
\def\iii{i}\def\plus{\raisebox{.5pt}{\tiny$+$\smpc}}

\def\spc{\hspace{1pt}}

\def\nspc{{\hspace{-2pt}}}
\def\ff{\mathsf\smpc f\smpc} 
\def\fff{\mbox{Y}}
\def\ww{{\mathsf w}}
\def\smpc{{\hspace{.5pt}}}

\def\zz{{\spc \mathsf z}}
\def\xx{{\mathsf x\smpc}}
\def\xxi{\mbox{\footnotesize \spc ${e^{S_0}}$}}
\def\jj{{\mathsf j}}
 \addtolength{\baselineskip}{.1mm}

\def\calO{{b}}
\def\be{\begin{equation}}
\def\ee{\end{equation}}

\def\bfU{\mbox{\textbf{\textit U}}}
\def\bfR{\mbox{\textbf{\textit R}}} 
\def\bfC{\mbox{\textbf{\textit C}}} 
\def\bfT{\mbox{\textbf{\textit T}}} 
\def\bbbi{{\, {{i}}\, }}
\def\nnn{n}

\def\spcm{\hspace{.25pt}}
\def\mathbi#1{\textbf{\em #1}} 
\def\som{{ \textit{\textbf s}}} 
\def\tom{{ \textit{\textbf t}}} 
\def\nnn{n} 
\def\mom{m} 
\def\la{\langle}
\def\bea{\begin{eqnarray}}
\def\eea{\end{eqnarray}}
\def\is{\!  & \!  = \!  &  \!}
\def\ra{\rangle}
\def\half{{\textstyle{\frac 12}}}
\def\cL{{\cal L}}
\def\halfi{{\textstyle{\frac i 2}}}
\def\ba{\bea}
\def\ea{\eea}
\def\lb{\langle}
\def\rb{\rangle}
\newcommand{\rep}[1]{\mathbf{#1}}

\def\uU{\bfU}
\def\be{\bea}
\def\ee{\eea}
\def\delbar{\overline{\partial}}
\def\ra{{\mbox{\large $\rangle$}}}
\def\la{{\mbox{\large $\langle$}}}
\def\ccdot{\!\spc\cdot\!\spc}
\def\nspc{\!\spc}
\def\tr{{\rm tr}}
\def\li{|\spc}
\def\ri{|\spc}
\def\OA{O_{\!\spc A}} 
\def\OC{O_C} 
\def\OB{O_B} 
\def\tMD{\mbox{\fontsize{7}{7}$\rm\smpc TMD$}}

\def\tFD{\mbox{\fontsize{7}{7}$\rm\smpc TFD$}}
\def\Oomega{\mbox{\fontsize{7}{7}$\rm\smpc \Omega$}}

\def\hf{\textstyle \frac 1 2}

\def\smpt{\hspace{.9pt}}
\def\bfcdot{\raisebox{-1.5pt}{\bf \LARGE $\spc \cdot\spc $}}
\def\spc{\hspace{1pt}}
\def\is{\! &  \! = \! & \!}
\def\d{{\partial}}
\def\n{{\bf \widehat n}}
\def\k{{\bf k}}
\def\GO{{\cal O}}
\def\rmAD{{\mathsf{Rad}}} 
\def\bbb{}
\def\pp{{\mbox{\tiny$+$}}}
\def\mm{{\mbox{\tiny$-$}}}
\def\ppp{\mbox{\fontsize{9}{8}$p$}}
\def\qqq{\mbox{\fontsize{9}{8}$q$}}
\def\TFD{\mbox{\fontsize{11}{13}$\rm\smpc TFD$}}
\setcounter{tocdepth}{2}
\renewcommand{\Large}{\large}
\addtolength{\baselineskip}{0.65mm}
\addtolength{\parskip}{.9mm}
\def\calO{{b}}
\def\be{\begin{equation}}
\def\ee{\end{equation}}
\def\half{\raisebox{-1pt}{\large $\frac 1 2$}}
\addtolength{\abovedisplayskip}{1.65mm}
\addtolength{\belowdisplayskip}{1.65mm}
\def\zzz{{\mbox{\large $z$\nspc\smpc}}}
\begin{titlepage}

\setcounter{page}{1} \baselineskip=15.5pt \thispagestyle{empty}

\vfil

${}$
\vspace{1cm}

\begin{center}
\def\thefootnote{\fnsymbol{footnote}}
\begin{changemargin}{0.05cm}{0.05cm} 
\begin{center}
{\Large \bf  Wormholes  in Quantum Mechanics}

\end{center} 
\end{changemargin}

~\\[1cm]
{Herman Verlinde\footnote{\href{mailto:verlinde@princeton.edu}{\protect\path{verlinde@princeton.edu}}}}
\\[0.3cm]

{\normalsize { \sl Physics Department,  
Princeton University, Princeton, NJ 08544, USA}} \\[3mm]

\end{center}

\begin{changemargin}{01cm}{1cm} 
{\small  \noindent 
\begin{center} 
\textbf{Abstract}\\[4mm]
\parbox{15 cm}{We introduce a geometric path integral definition of wormhole partition functions in a general class of 1D quantum systems obtained by quantizing a phase space. We compute the wormhole partition function in a semi-classical limit and in some simple examples. The partition function of the $n$-fold wormhole is found to be identical to the $n$-th R\'enyi entropy of a thermal mixed state of the doubled system.
This mixed state incorporates three types of quantum statistical behavior: classically correlated, quantum entangled, and classically uncorrelated. 
We apply our prescription to 2D CFTs with Virasoro symmetry and recover the holographic dual formulation in terms of AdS${}_3$ gravity.  }

\end{center} }

\end{changemargin}
 \vspace{0.3cm}
\vfil
\begin{flushleft}

\today
\end{flushleft}

\end{titlepage}

\flushbottom

\newpage
\tableofcontents
\thispagestyle{empty}

\setcounter{page}{0}
\newpage

\setcounter{footnote}{0}

\section{Introduction}
\vspace{-2mm}

\vspace{-1mm}

Wormhole contributions to the gravitational path-integral play a key role in the recent studies of the quantum black holes and their holographic duals~\cite{replica1}-\cite{Saad:2019}. The wormhole saddle points are essential for explaining, among others,  the late-time behavior of the spectral form factor and the Page curve in models of evaporating black holes \cite{Cotler:2016}-\cite{Almheiri:2019b}.  

At first sight, the wormholes that explain the Page curve look different from those that contribute to the spectral form factor.  The replica wormholes capture aspects of decoherence and purification of black hole radiation \cite{Penington:2019}-\cite{Almheiri:2019b}, while the wormhole contributions to the form factor quantify dephasing between two decoupled systems \cite{Saad:2018}. In both contexts, it is usually implied that wormhole saddle points are a characteristic of gravitational systems and their holographic duals \cite{Witten:1999}-\cite{Stanford:2020}.

In this paper we will show that wormhole geometries have a natural place in a large class of ordinary quantum systems, ranging from systems as simple as a coupled oscillator or a particle moving on a symmetric space to 2D CFTs with Virasoro symmetry. The basic idea is as follows. Consider a quantum system obtained by quantizing a phase space parametrized by generalized coordinates and momenta $X^a$, symplectic form $\Omega = \frac 1 2 \omega_{ab}\spc dX^a\!\wedge dX^b$ and Hamiltonian $H$. The thermal partition function $Z(\beta) \nspc =\nspc {\rm tr}(e^{-\beta H})$ of this quantum system has the following path integral representation \cite{Kontsevich:1997}-\cite{Verlinde:1989} 
\bea
\label{zex}
Z(\beta) \is \int [dX] \, e^{\, \mbox{\footnotesize $
{\int_D\nspc \Omega - \oint_{\partial D} \nspc Hdt}$}}
\eea
where $D$ denotes the disk with the thermal circle as boundary. 

If the symplectic form is exact, $\Omega = d\alpha$,  we can write $\int_D\Omega = \oint_{\partial D} \alpha$ and the above formula can be reduced to the usual 1D path integral formula. However, there are many interesting systems for which $\Omega$ is closed but not exact. For such systems the formula \eqref{zex} identifies  the thermal partition function $Z(\beta)$ with the functional integral  $Z(D)$ associated with a disk 
 \bea
 \label{zdisk}
Z(\beta) \is 
Z(D)  \qquad ; \qquad D\, =\;\; \raisebox{-5mm}{$ \includegraphics[scale=.38]{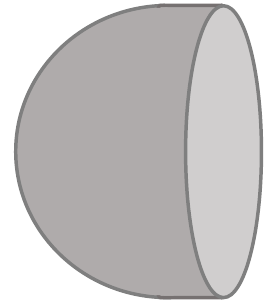}$}.
\eea
In systems with gravity duals, the holographic dictionary  identifies the disk $D$ with the longitudinal $(r,t)$ plane of a euclidean black hole solution with the horizon at the center. In our simpler setting, $D$ is just used as an auxiliary space to write the action integral.  
Since the symplectic form $\Omega$ is closed, the integrand in \eqref{zex} is topological in the bulk and only depends on the boundary values of $X$. The local geometric properties of $D$ therefore do not matter. However, its topology does matter.

It is now clear how we can introduce wormhole geometries for this class of quantum systems. Let $\Sigma$ denote some general (connected) two-dimensional surface with one or more boundary components $\partial \Sigma$. We then associate a partition function to  $\Sigma$ by means of the following path integral formula \cite{Verlinde:1989}-\cite{Hirshfeld:1999}
\bea
\label{zsigm}
Z(\Sigma) \is \int [dX] \, e^{\, \mbox{\footnotesize $
{\int_\Sigma\nspc \Omega - \oint_{\partial \Sigma} \nspc Hdt}$}}
\eea
Here $t$ denotes some preferred (euclidean) time coordinate on the boundary $\partial \Sigma$.  The integrand of the partition function \eqref{zsigm} only depends on the boundary values of $X$. However, we will see that in the general class of systems considered below, the partition function $Z(\Sigma)$ does not factorize into a product of partition functions associated to each boundary.

 Let $\Sigma_n$ denote the $n$-fold trumpet geometry with $n$ thermal circles as its boundary, all with the same boundary length equal to some inverse temperature $\beta$. Then the partition function associated with $\Sigma_n$ in general yields a non-factorized result
\bea
\label{ntrumpet}
\ \ \ \la Z(\beta)^n \ra  \, = \,  {Z(\Sigma_n)}\qquad & & \qquad \Sigma_n \, = \,  \raisebox{-1cm}{$ \includegraphics[scale=.48]{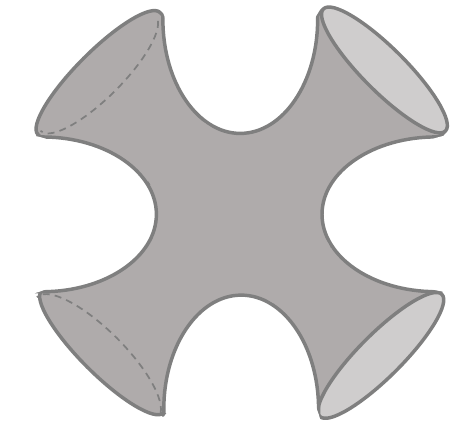}$}. 
\eea

We will compute this partition function for a general class of quantum systems. We will do this via two methods. The first is through direct evaluation of the path integral. A second  indirect method is to first demonstrate that the partition function \eqref{ntrumpet} represents the  $n$-th R\'enyi entropy of a suitable thermal mixed state of the two-sided system. We will call this density matrix the {\it thermo-mixed double}~\cite{HV:2020}-\cite{Takayanagi:2020} and refer to the relation between its R\'enyi entropy and the wormhole partition function 
\bea
\label{tmddef}
{\rm tr}(\rho_{\tMD}^n) \, = \, \frac{{\la Z(\beta)^n \ra}{\strut}}{Z(\beta)^n{\strut}} \, =\,  \frac{ {Z(\Sigma_n)}}{Z(D)^n}
\eea
as the {\it replica Ansatz}. The self-consistency of this Ansatz provides an independent check of our geometric prescription.

In section 2 we describe the class of quantum systems and define the partition function \eqref{zsigm}. In section 3, we compute $Z(\Sigma_n)$ via a general semi-classical path integral computation and through an exact computation for the particle on a group manifold. 
In section 4, we postulate an explicit form of the thermal mixed double state $\rho_{\tMD}$ for the systems of interest and match its $n$-th R\'enyi entropy with the $n$ wormhole partition function $Z(\Sigma_n)$.  In the Appendix we discuss the application of our prescription to the coupled oscillator, Schwarzian quantum mechanics  and 2D CFTs with Virasoro symmetry \cite{Iliesiu:2019}-\cite{Cotler:2018}.

\section{Wormholes in Quantum Mechanics}

\vspace{-1mm}

We consider the general class of quantum systems obtained by quantizing a classical phase space parametrized by generalized coordinates and momenta $(\tau^I, X^a)$ with symplectic form and Hamiltonian of the form
\bea
\label{ohm}
\Omega \is  
 \frac 1 2\, 
\omega_{a \bbb{b}} \, dX^a\! \wedge dX^{\smpc b}  + \spc dJ_I \nspc \wedge\nspc\smpc d \tau^{\smpc I}\spc,\qquad \qquad
H \spc = \spc H_0(J) \smpc + \smpc h(X).  
\label{ham}
\eea
The relevant non-zero Poisson brackets associated with $\Omega$ are
\bea
\label{poissonb}
 \{ X^a,X^b\} =  \omega^{ab} \quad & & \quad \{J_I,\tau^J\} = \delta_{I}{}^J , 
\eea
where $\omega^{ab}$ is the inverse of the matrix $\omega_{\spc ab}$.
The $J_I$ denote a set of  central functions $J_I(X)$ with the property that their Poisson bracket with any of the $X^a$ variables identically vanishes  
\bea
\{X^a, J_I \} \!\is\!  \omega^{\spc ab} \partial_a J_I \spc = \, 0.
\eea
The Hamiltonian \eqref{ham} does not depend on the $\tau^I$ variables. These properties identify $J_I(X)$ as a set of conserved quantities, and ($J_I,\tau^I$) as a corrresponding set of action-angle variables.  Appendix A describes three concrete simple examples systems of this type.

  Let $\Sigma$ denotes some two dimensional surface with a one dimensional boundary $\partial \Sigma$ parametrized by a (euclidean) time coordinate $t$. Without loss of generality, we can assume that $\Sigma$ is connected; the boundary $\partial \Sigma$ may have multiple components.  We can associate a partition function $Z(\Sigma)$ to this surface $\Sigma$ via the following path integral formula
\bea
\label{zsigma}
 && \hspace{-4mm} Z(\Sigma)   = \,   \int \! \spc [dX d\tau] \spc \exp\Bigl(\spc {\mbox{\Large $\frac{i}{\hbar}$} \spc S_{\Sigma}[X ]}\spc \Bigr)\\[4mm]
S_{\Sigma}[X,\tau]\, \is   \int_\Sigma\nspc \bigl(\mbox{\Large $\frac 1 2$}  \spc \omega_{ab}\, dX^a \! \wedge dX^b + \spc dJ_I \nspc \wedge\nspc\smpc d \tau^{\smpc I}\bigr) \spc  - \spc \oint_{\partial \Sigma} \!\!  H dt \spc 
 \eea
Here $X$ and $\tau$ are 2D fields defined on the surface $\Sigma$. The boundary values of $\tau$ are integrated over, while the $X$ fields satisfy either Neumann (for partition functions) or Dirichlet (for wave functionals) boundary conditions at $\partial \Sigma$. Note, however, that the 2D part of the action functional is an integral of a closed two-form and is therefore topological: $S[X]$ is a unique functional of the boundary values of $X$, up to integer multiples of  $2\pi i \hbar$. This property ensures that the functional integral \eqref{zsigma} is well defined.  In case $\Sigma$ is a disk, the path integral \eqref{zsigma} defines the thermal partition function. In the following, we will consider more general $n$-fold trumpet geometries with multiple boundaries.

The  $\tau_I$ act as a lagrange multipliers that impose the conservation condition $dJ^I=0$ along each boundary component of $\Sigma$.  The allowed  spectrum of $J^I$ follows from the geometric quantization condition that the integral of the symplectic form $\Omega$ over any 2-cycle (= quantizable orbit) must be an integer multiple of $2\pi i \hbar$. After reduction to an orbit with fixed $J_I$, the symplectic form simplifies to $\omega  = \frac 1 2 \, \omega_{ab} \spc dX^a\!\!\smpc \wedge \!\smpc d X^b .$ This reduced form is still closed, $d\omega=0$, but in general not exact $\omega \neq d \alpha$. Hence the integral $\int_\Sigma \omega$ over an open phase space surface $\Sigma$ with boundary $\partial \Sigma$ can not be written as an integral over the boundary. This non-exactness of the symplectic form is a generic phenomenon in quantum many body systems, and leads to the presence of non-integrable Berry-Aharanov-Bohm phases. Specific examples of dynamical systems of this type are the particles on a group manifold and CFTs with Kac-Moody or Virasoro symmetry.

The above description of the classical phase space motivates to the following terminology for the corresponding quantum mechanical operators
\bea
{\cal O}(\tau,X)  \ \supset \  \left\{ \begin{array}{ccc} {J_I(X)} \is {\rm Casimir \ operators} \\[1.3mm]
{X}^a  \is {\rm currents}\\[1.3mm]
{W(\tau)} \is {\rm open \ Wilson\ lines} \end{array} \right. .
\eea
The $X^a$  commute with the unperturbed Hamiltonian $H_0$ in eqn \eqref{ohm}, and thus generate a symmetry algebra of $H_0$. The $X^a$  and $J_I$ satisfy the commutation relations $\[ X^a,X^c  \] = \omega^{\spc ac}$ and $\[J_I, X^a \] \, = \, 0.$
Acting with the $X^a$  on an eigenstate of the Casimirs $J_I$ produces a module, or irreducible representation of the $X^a$ algebra. 
The Wilson line type operators $W(\tau)$ do not commute with the Casimirs and interpolate between different eigensectors. 

The Hilbert space decomposes into a direct sum of subsectors ${\cal H}_{j}$, each spanned by an orthonormal basis of $J_I$  eigen states 
\bea
J_I |jms\rangle = j_I |j ms\rangle , \qquad \qquad X^a |j ms\rangle =  \sum_{m'} \, |j m's\rangle \langle jm' |X^a |j m\rangle
\eea  
The quantum number $s$ labels separate but otherwise identical irreducible representations of the $X$ operator algebra with the same value of the Casimir $J_I$. We will denote the degeneracy by $d_j$; so $s$ ranges from $1$ to $d_j$.
Thanks to the decomposition \eqref{ham} of the Hamiltonian, we can label the energy eigenstates and eigenvalues as
\bea
\label{esplit}
H |jms\rangle \spc =\spc E_{jm} |jms\rangle , \qquad \qquad E_{jm} \is E_j+ e_{jm}
\eea
with $E_j$ the eigen value of the unperturbed Hamiltonian $H_0(J)$  and $e_{jm}$ the eigenvalue of the second term $h(X)$. The $E_j$ contribution is constant within a given module, while $e_{jm}$ encodes the energy carried by the currents $X^a$.  Each energy level $E_{jm}$ has degeneracy $d_j$, labeled by $s$. In CFT language, $E_j$ equals the conformal dimension of the primary state, and $e_{jm}$ indicates the energy gap between the primary and the  descendent states.

For all quantum mechanical systems that can be organized in the above fashion, we can associate a partition function $Z(\Sigma)$ to any closed surface $\Sigma$ with or without boundaries. If we restrict our attention to only the Casimir and current observables, we are free to integrate out the $\tau_I$ variables. The remaining functional integral can be recast in first order form   \cite{Schaller:1994}
\bea
\label{zsigmanew}
& &\!\!\!\! \hspace{-3mm} Z(\Sigma)  = \int \! \spc [dX d\eta\spc\spc] \,\spc e^{\mbox{\footnotesize $\frac{i}{\hbar} \smpc S_{\Sigma}[X,\eta\spc ]$}}\\[4mm]
S_{\Sigma}[X,\eta] \is \int_\Sigma \nspc
( \eta_a \!\wedge dX^a +\frac 1 2 \omega^{\spc ab} \eta_a\!\wedge \eta_b )  
 \,   
-\, \oint_{\partial \Sigma} \!  H dt \, 
\label{etact}
 \eea
 Here we introduced a set of one-form variables $\eta_a$ dual to the scalar variables $X^a$. Performing the gaussian integral over the one form variables $\eta_a$ produces the original expression \eqref{zsigma} of the partition function~$Z[\Sigma]$, with the properly defined functional measure. As we will see shortly, the above first order form is more convenient for performing explicit computations. For the special case of the particle on a group manifold, the $X^a$ form the generators of a Lie algebra and the action \eqref{etact} takes the form of a BF gauge theory. More generally, it defines a so-called Poisson sigma model associated with the Poisson algebra $\{X^a,X^b\} = \omega^{\spc ab}$. 

The functional integral  \eqref{zsigmanew} on surfaces $\Sigma$ with boundaries is performed with Neumann boundary conditions on the fields $X^a$.
We can generalize the functional integral prescription to define quantum states, or density matrices, by imposing Dirichlet boundary conditions on the fields $X^a$ and fix the value of  $X^a(s)$ along one or more of the boundary segments of $\Sigma$. (Section 4.1 describes three examples.) The resulting functional integral defines a wave-functional $ \Psi[X]$ of the field $X^a(s)$, satisfying the local constraint equation  \cite{Schaller:1994}\cite{Verlinde:1989}
\bea
\label{psidef}
\Bigl(\omega^{ab} \frac{\delta\ }{\delta X^b} - \frac{dX^a\!\!}{ds} \;\spc\Bigr) \Psi[X] \is 0 
\eea
This constraint is analogous to the Gauss law constraint in gauge theory: it is the quantum manifestation of the time component of the classical equation of motion $dX^a = \omega^{ab} \eta_b$, obtained by varying the one form field $\eta_a$ in the action \eqref{zsigmanew}.  It uniquely fixes the local dependence of $\Psi[X]$ on the path $X(s)$. The inner product is defined by multiplying the two wave functionals and performing the (properly gauge fixed) functional integral over $X(s)$.

\section{Wormhole Partition Function}
\vspace{-1mm}

In this section we will compute the partition function $Z(\Sigma_n)$ associated with the $n$-fold  wormhole geometry $\Sigma_n$ depicted in equation \eqref{ntrumpet}. 
Let us already announce the result. 

Due to the form \eqref{esplit} of the energy spectrum, the thermal partition function $Z(\beta)$ splits up as a sum
\bea
\label{zbetaj}
Z(\beta) \is 
\sum_j \spc 
Z_j(\beta) \qquad \qquad Z_j(\beta) \, \equiv \, \sum_{m,s} e^{-\beta \spc E_{jm}} \, \equiv \, d_j\spc e^{-\beta E_j} \spc  \zzz_j(\beta)
\eea
of thermal partition functions at fixed eigenvalue for the Casimirs $J_I$. Here $\zzz_j(\beta) = \sum_m e^{-\beta e_{jm}}$. With this notation, the result for general partition function $Z(\Sigma_n)$ reads 
\bea
Z(\Sigma_n) \is \nspc \sum_j \, Z_j(\Sigma_n) \qquad \quad \ \  \ Z_j(\Sigma_n) \spc=\spc  \frac {Z_j(\beta)^n} {d^{2n-2}_j\!\!} \;  \spc = \spc  \frac{ e^{-n\beta E_j} \zzz_j(\beta)^n} {d^{n-2}_j\!\!}. \; 
\label{ztwo}
\eea
In the next section, we will show that this result is equal the $n$-th R\'enyi entropy of a specific mixed state $\rho_{\tMD}$. Here we will verify this expression via a semi-classical evaluation of the functional integral definition of $Z_j(\Sigma_n)$ and through an exact non-perturbative computation for the special case of the particle on a group manifold.

\subsection{Semi-classical computation}
\vspace{-1mm}

The key observation that will help guide our calculations is that the expectation value, defined via the functional integral \eqref{zsigma}, of one of the Casimir operators inserted at a bulk point $P$ of $\Sigma$ does not depend on the position of the point $P$. This result, first proved by Cattaneo and Felder, is plausible, given that all Casimir operators are conserved quantities and central functions with vanishing Poisson bracket with all $X^a$ variables \cite{Cattaneo:1999}. Note that we can choose this point $P$ to approach any of the boundary components of $\partial \Sigma$.
 Combining the above observations, we conclude that the value of the Casimir $J_I$ takes the same constant value along all boundary components of $\partial \Sigma$. This constraint is the key physical property of $Z(\Sigma)$ that follows from the connectedness of the surface $\Sigma$.

We temporarily set the extra Hamiltonian $h(X)$ in equation \eqref{ohm} equal to zero, keeping only the unperturbed Hamiltonian $H_0(J)$ with eigen energies $E_j$. The action \eqref{zsigma} then reduces to that of a 2D Poisson sigma model with a boundary term
\bea
\label{spoisson}
 S_\Sigma[\eta,X] \is \int_{\smpc \Sigma} \nspc
( \eta_a \!\wedge dX^a + \omega^{\spc ab} \eta_a\!\wedge \eta_b ) - \oint H_0(X) dt \eea
The Hamiltonian $H_0(X)$ is a central function with the property $\omega^{\spc ab}\partial_b H_0 =0$.
The bulk action does not depend on a choice of metric on $\Sigma$, and is therefore `topological', or diffeomorphism invariant. The sigma-model action \eqref{spoisson} is invariant under the gauge symmetry 
  \bea
  \label{gaugetrafo}
 \delta_\epsilon X^a \! \is\nspc \omega^{\spc ab} \epsilon_b\qquad \qquad \delta_\epsilon \eta_a \spc= \spc 
 d\epsilon_a + \partial_a\omega^{\; bc}\, \eta_b \, \epsilon_c.
 \eea 
 
The partition function and the expectation values of operators of Poisson sigma models on the disk are well studied \cite{Cattaneo:1999}-\cite{Hirshfeld:1999}, and form a key tool in the physical proof of the general formula for the deformation quantization of a general Poisson manifold \cite{Kontsevich:1997}. One of the key results of \cite{Cattaneo:1999} is that the functional integral $Z(\Sigma)$ decomposes in to a sum of restricted functional integrals $Z_j(\Sigma)$ with fixed eigen values $j_I$ of the Casimir functions $J_I(X)$, as in the left equation in \eqref{ztwo}.  Below we will describe the semi-classical computation of the restricted partition function $Z_j(\Sigma_n)$. The restricted phase space for fixed $j$ will be denoted by ${\cal S}_j$.  In the special case that $X^a$ generate a Lie algebra, the partition functions $Z_j(\Sigma_n)$ can be computed exactly \cite{Blau:1993}. In general, the symplectic leaves ${\cal S}_j$ can be complicated manifolds and it will not be easy to explicitly evaluate its partition function in the most general setting. 

We now make two simplifying assumptions: 1) we work in the semi-classical  limit $\hbar \to 0$, 2) we assume that ${\cal S}_j$ is a homogenous space. For small $\hbar$, the dimension of the Hilbert space is equal to the phase space volume in units of $\hbar$. So we can replace our prediction \eqref{ztwo} by the geometrical formula $Z_j(\Sigma_n) ={\rm Vol}({\cal S}_j)^{2-n}$.
We wish to derive this result by performing the functional integral using a semi-classical expansion $
X^a = \bar{X}^a \, +\spc \delta\nspc\smpc X^a$  around a given classical solution $X^a_{\rm cl} = \bar{X}{}^a$, while 
keeping only the terms quadratic in $\delta X^a$ and $\eta_a$.  In case ${\cal S}_j$ is a homogenous manifold, we can use  the Pfaffian of the symplectic form $\bar{\omega}^{\spc ab} = \omega(\bar{X})^{ab}$ evaluated at $\bar{X}^a$ as a local proxy for the volume of ${\cal S}_j$ \cite{Hirshfeld:1999}. 

Via this reasoning, we arrive at the following prediction for the partition function at fixed values for the Casimirs 
 \bea
 \label{zpredict}
Z_j(\Sigma_n) \is\nspc d_j^{\spc 2-n} \spc = \spc {\rm Vol}({\cal S}_j)^{2-n} = {\rm Pf}(\bar\omega)^{2-n}
 \eea
We now outline the derivation of this result, following \cite{Hirshfeld:1999}. To fix the gauge invariance \eqref{gaugetrafo}, we impose the gauge condition
 \bea
 \label{gf}
 \chi_a  \equiv   
 d^{\spc\star} \eta_a + \spc g_{ab} \spc \delta\nspc\smpc X^b\! \is\nspc 0\, ,
 \eea
with $d^{\spc \star} = \star\spc d\spc \star$ and $g_{ab}$ some arbitrarily chosen metric. As we will see, the second term in \eqref{gf} is required to eliminate a residual zero mode in the first term, while the first term is needed to produce a well behaved ghost kinetic term. 

Following the standard Faddeev-Popov gauge fixing procedure, we add the following two terms to the action $
 {S}_{\rm gh} + {S}_{\rm gf} = \int_{\smpc \Sigma}( \bar{c}_a (g^{ab} \Delta + \; \overline{\!\omega\!}\,{}^{ab}  )c_b + \lambda{}^a (  d^{\spc \star} \eta_a
 + \spc g_{ab}\spc \delta\nspc\smpc X^b))$,
with $\Delta = d^{\spc \star} d$ the scalar laplacian on $\Sigma$. Performing the gaussian integral over $\delta X^a$ produces the total action
\bea
\label{quad}
{S} \is \int_\Sigma\nspc \bigl(   g^{ab} d^{\spc *}  \eta_a d\eta_b 
 +\spc \overline{\!\omega\!}\,{}^{ab}
\eta_a \! \wedge \nspc \eta_b  + \bar{c}_a (g^{ab} \Delta + \; \overline{\!\omega\!}\,{}^{ab}  )c_b\bigr)
\eea
This is a quadratic action, so we can just perform the remaining gaussian functional integral. It is useful to first make a further simplification. The action \eqref{quad} is invariant under the nilpotent BRST symmetry $
\delta_{\raisebox{-1pt}{\tiny Q}} \eta_a =  d c_a ,$ 
$ \delta_{\raisebox{-1pt}{\tiny Q}}  \bar{c}_a = d^{\spc \star} \eta_a$. The change of the lagrangian under a small variation $\delta g_{ab}$ of the metric can be written as a total BRST variation
$\delta g^{ab} (d^{\spc *}  \eta_a d \eta_b 
 +\spc \bar{c}_a \Delta c_b)
= \delta_{\raisebox{-1pt}{\tiny Q}} \bigl(\delta g^{ab} \bar{c}_a d \eta_b 
 \bigr)$.
Since this variation leaves all physical quantities invariant, we are free to take the limit in which inverse metric $g^{ab}$ becomes infinite. The functional integral then localizes on the space of zero modes $d\eta_a = d^{\spc *}  \eta_a =0,$ and $\Delta c_a  = \Delta \bar{c}_a =0. $

On the $n$-fold replica wormhole geometry $\Sigma_n$, there are $n$ harmonic one-forms $\omega_i$, one for each boundary component. Correspondingly,  we can decompose the harmonic one form field as a sum over zero modes $\eta_a = \sum_i \eta_a^i \alpha_i$, with $\alpha_i$ an orthonormal basis of harmonic one forms. There is only one constant zero mode for each scalar field $\bar{c}_a$ and $c_a$. Hence, in the limit that $g^{ab}$ becomes infinite, the functional integral reduces to a finite dimensional gaussian integral
\bea
\int_{\strut{\raisebox{-3pt}{\scriptsize \ zero modes}}}\hspace{-1.425 cm} d\eta  \spc d\bar{c}\spc  dc \;\, e^{\mbox{\footnotesize $- \int_\Sigma (\bar\omega^{\smpc ab}  \eta_a\!\wedge\nspc \spc \eta_{b}  + \bar\omega^{\smpc ab} \bar{c}_a c_b)$}}\is \frac{\; ({\rm Pf}(\bar\omega_{\smpc 0}))^2} {({{\rm Pf} (\bar\omega_{1}))^{n}\!\!}}
\label{semisimple}
\eea
The numerator is the contribution of the integral over the constant fermionic zero modes of the ghost fields $\bar{c}_a$ and $c_a$, and the denomitor is the contribution of the gaussian integral over the harmonic modes of the one-form field $\eta_a$. Correspondingly, $\bar\omega_0^{ab}$ denotes the integral of $\omega^{ab}(\bar{X})$, and $\bar\omega_1^{ab}$ denotes the integral of $\omega^{ab}(\bar{X})$ against the wedge product of two normalized harmonic one forms $\alpha_i$. If $\bar{X} = X_{\rm cl}$ is a constant, then $\omega_0^{ab} = \omega_1^{ab}$ and the above formula reproduces our semi-classical prediction \eqref{zpredict} for the partition function $Z_j(\Sigma_n)$.  

We leave the study of the general case with $h(X)$ non-zero for the future; here we only make a brief qualitative comment. Imagine turning on $h(X)$ via a perturbative expansion. For this it will be sufficient to compute correlation functions of an arbitrary number of insertions of $h(X)$ on the boundaries of $\Sigma_n$.  Work by Kontsevich and others has shown that functions of $X$ inserted on the boundary of a disk generate an operator algebra isomorphic to the star product algebra of functions on the Poisson manifold \cite{Kontsevich:1997}\cite{Cattaneo:1999}. The partition function $\zzz_j(\beta)$ can  be thought of as the trace of the operator $e^{-\beta h(X)}$ over the representation of the star product algebra labeled by the Casimir eigenvalues $j_I$. It seems reasonable to assume that on a surface $\Sigma_n$ with several boundaries,  insertions of operators on different boundaries generate separate mutually commuting star product algebras. The partition function at fixed $j$ thus naturally contains a factor $\zzz_j(\beta)^n$, reproducing the anticipated result \eqref{ztwo}. We will now make this argument more precise for the case of the particle on a group.

 \subsection{Particle on a group I}
 \vspace{-1mm}

A concrete example of the above class of quantum systems is the particle moving on a group manifold $G$ with a lagrangian of the following form
$L\! = \frac 1 2 \, {\rm tr}((g^{-1} (\partial_t \!- A) g)^2)$
with $g(t) \in G$.  Here $A \in \mathfrak{g}$ denotes a constant Lie algebra element. The corresponding first order action reads
\bea
\label{gactt}
S&\!\!\! =&\!\!\! \int\! dt \spc \bigl( {\rm tr}(\lambda (g^{-1}\nspc \partial_t g)) -  H\bigr) , \qquad \qquad H =\half\, {\rm tr}( \lambda^2)\spc +\spc  {\rm tr}(A\spc g \lambda \smpc g^{-1})\, ,
\eea
where we introduced the momentum variable given by the dynamical Lie algebra element $\lambda(t) \in \mathfrak{g}$. For constant $\lambda$, the kinetic term in \eqref{gactt} defines the familiar coadjoint orbit action of the group $G$. From the corresponding Poisson brackets, one derives that the variables
\bea
X^a \is {\rm tr}(\tau^a g\lambda g^{-1}), \qquad \qquad J_I(X) = {\rm Casimir\ functions}
\eea
upon quantization satisfy the Lie algebra commutation relations $[X^a,X^b] = f^{ab}{\!}_c X^c$, while the Casimirs are central functions with the property $[J_I,X^a] =0$. The Hilbert space of the particle on the group manifold is given by the space ${\cal H}$ = Fun$(G)$ of functions on $G$. This space decomposes into a sum over all irreducible representation $R_j$ of $G$, labeled by the eigenvalues $j_I$ of the Casimir operators $J_I$. Each representation $R_j$ appears with multiplicity $d_j =$ dim($R_j$). So the structure of the Hilbert space fits the pattern described above.

The group acts on the Hilbert space via unitary operators $U(g) = \oplus_{j,s} U_{R^{(s)}_j}(g)$.  Hence if ${\rm tr}_{\cal H}$ denotes the trace over the full Hilbert space and $\chi_j(g) = {\rm tr}_{R_j}(U_{R_j}(g))$  the character of the representation $R_j$, we have ${\rm tr}_{\cal H}(U(g)) = \sum_j d_j \chi_j(g)$. When acting on a given representation $R_j$, we have $e^{-\oint dt H} = e^{-\beta E_j} U_{R_j}(g_\beta)$ with $g_\beta = e^{-\beta A^a \tau_a}$. We thus find that the partition function of the particle on the group reads 
\bea
\label{zgbeta}
Z(\beta)\nspc \is\nspc \sum_j \, d_j\,  \chi_j(g_\beta) \, e^{-\beta E_j}
\eea
As shown in Appendix A2, the $n$-fold wormhole  partition function $Z(\Sigma_n)$ reads \cite{Iliesiu:2019b}
\bea
\label{zcheckt}
Z(\Sigma_n) \nspc \is \nspc \sum_j \, \frac {e^{-n\beta E_j}\spc \chi_j(g_\beta)^n} {d_j^{n-2}\!}
\eea
This matches with the announced result \eqref{zpredict}  via the identification of $\chi_j(g_\beta)= \zzz_j(\beta)$.

\smallskip

\section{A New Thermal Mixed State}

\vspace{-1mm}

In this section we will use the  results of the previous section to motivate the definition of a new type of thermal state. We will specify this state both geometrically and through its quantum statistical properties. By construction, the $n$-th R\'enyi entropy of the thermo-mixed double state reproduces the partition function $Z(\Sigma_n)$ associated with the $n$-fold replica wormhole geometry. We quantify and compare its entanglement, von Neumann and R\'enyi entropy with other standard thermal states.

\subsection{Three thermal states }

\vspace{-1mm}

 Using our path integral definition \eqref{zsigmanew}, we define three thermal states: the thermal density matrix $\rho_R$, the thermofield double $|\TFD\ra$ and the thermo-mixed double $\rho_{\tMD}$
\bea
\label{holothermr}
\ \rho_R \is   \frac{Z(\Sigma_R)}{\; Z(\beta)}\qquad \ \, ; \qquad |\TFD \ra\, = \,  \frac{Z(\Sigma_{\tFD})}{\; \sqrt{Z(\beta)}}  \qquad \, ; \qquad \rho_{\tMD} \, = \, \frac{Z(\Sigma_{\tMD})}{\; Z(\beta)}.\qquad
\\[6mm]\label{holotherms}
\Sigma_R \is \raisebox{-6.5mm}{$ \includegraphics[scale=.49]{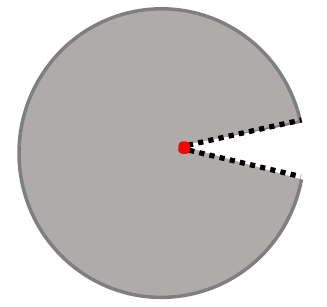}$}  \quad\ \ ; \qquad\; \;\, \Sigma_{\tFD} = \, \raisebox{-3.7mm}{$  \includegraphics[scale=.49]{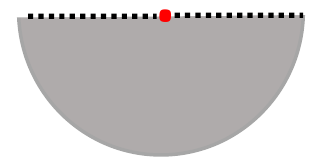} $}  \qquad\, ; \qquad \Sigma_{\tMD} \, = \, \raisebox{-6.4mm}{$ \includegraphics[scale=.5]{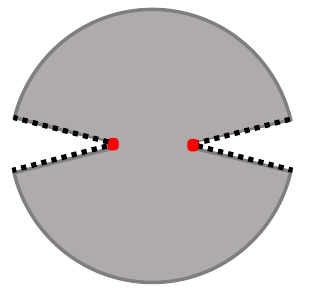}$}.\qquad
\eea
Here the circular boundary denotes the thermal circle (left), the thermal half-circle (middle) and two thermal half-circles (right).  The filled-in disk or half-disks represent the corresponding euclidean bulk space-time region over which we integrate the symplectic two-form~$\Omega$.  

Topologically, we can view these 2D space-times as a segment of a euclidean 2D black hole space time. The dashed straight segments depict the bulk state, given by functionals of the boundary values $X^a(s)$ of the bulk fields, supported within their respective entanglement wedges. As explained in \cite{HV:2020}, the above geometric definition \eqref{holothermr} of the TMD state can readily be seen to be equivalent to the replica wormhole definition \eqref{tmddef} of its R\'enyi entropies: gluing together $n$ copies of $\Sigma_{\tMD}$ according to the replica method for computing the $n$ fold product $\rho_{\tMD}^n$, produces the $n$ fold replica wormhole geometry $\Sigma_n$. 
 For the specific case of the double trumpet partition function $Z(\Sigma_2)$, the relationship \eqref{tmddef} with the TMD state is geometrically represented as 
\bea
\label{tmdglue}
\raisebox{-8mm}{$\includegraphics[scale=.53]{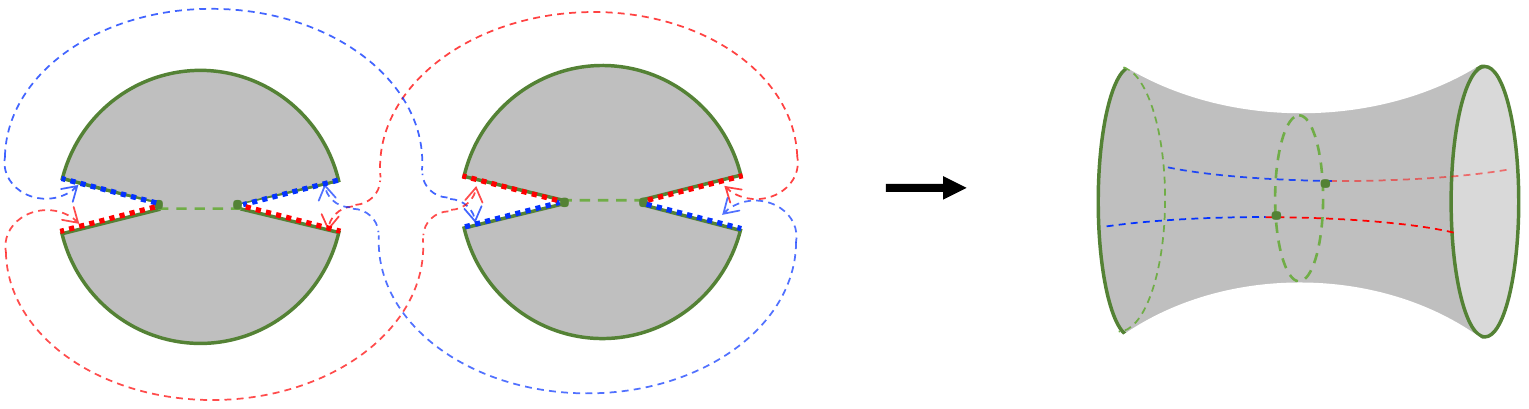}$}
\eea
We will determine the explicit form of $\rho_{\tMD}$ momentarily, using this geometric path-integral prescription and the result \eqref{ztwo} for the partition function $Z(\Sigma_n)$ as our guide.

As a warm-up, we write the first two thermal states for our specific quantum system. The spectrum and thermal partition function are given in \eqref{esplit} and \eqref{zbetaj}. It will be useful to introduce the probability distribution on the space of Casimir eigenvalues 
\bea
\label{pj}
 p_j \is \frac{\spc Z_j(\beta)\spc }{Z(\beta)} \, = \, \frac{d_j \spc e^{-\beta E_j} \spc \zzz_j(\beta)\spc }{Z(\beta)}, 
\eea
with $Z_j(\beta)$ defined in \eqref{zbetaj} and $\zzz(\beta) = \sum_m e^{-\beta e_{jm}}$. The thermal density matrix $\rho_R$ then takes the form 
\bea
\label{rhorj}
 \rho_R\! \is\! 
 \sum_{j} \,  \frac{p_j}{d_j}
 \, \rho_{j}  \qquad \ \
\rho_{j} =\smpc
\sum_{m,s} \, q_{jm}  \, |jm s\rangle_R\langle jm s |, \qquad \  \ q_{jm} \spc = \, \frac{\!e^{-\beta \spc e_{jm}}\!\!}{\spc \zzz_j(\beta)}\, 
\eea
of a weighted sum  of thermal density matrices  $\rho_{j}$ within each sector labeled by $j$. 

Similarly, we can write the thermo-field double state as a coherent sum
\bea
\label{tfdj}
|\TFD\rangle\! \is\! 
\sum_{j} \sqrt{p_{j}\!}\;\spc |\TFD_j\rangle,\\[3mm]
|\TFD_j\rangle \! \is\! 
 \sum_{m,s} \spc \sqrt{q_{jm}\!}\;\spc \spc |jms\rangle_{\!\smpc L}\spc |jms\ra_{\nspc R}
\eea
This TFD state carries maximal entanglement but has zero von Neumann entropy. It still counts as a thermal state: the partial trace of $\rho_{\tFD} = |\TFD\ra\la\TFD|$ over the left Hilbert space produces the thermal density matrix $\rho_R$ on the right Hilbert space. 
The entanglement entropy encoded in the TFD state thus equals the thermal entropy of $\rho_R$. We will write the formulas for the entanglement and entropy of both states in the next subsection.

The TFD stands in contrast with the factorized thermal state
\bea\rho_L \otimes \rho_R  \is  \frac{Z(\Sigma_L \cup \Sigma_R)}{Z(\beta)^2},
 \qquad \qquad 
\Sigma_L \cup \Sigma_R\; =  \;  \raisebox{-6.5mm}{$ \includegraphics[scale=.46]{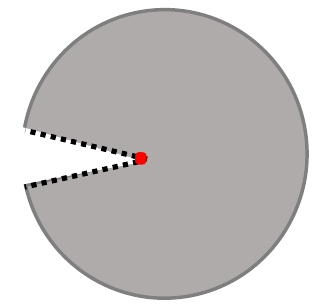}$} \ \raisebox{-6.75mm}{$ \includegraphics[scale=.465]{pacman.pdf}$}
\eea 
which has zero entanglement entropy between the two sides. In our geometric language, it is associated with a factorized geometry $\Sigma_L \cup \Sigma_R$. In the AdS/CFT context, the factorized thermal state describes a disconnected pair of black holes without an ER bridge and twice the BH entropy. It arises if one puts two CFTs in contact with two separate environments and lets them thermalize independently.

\medskip

With this bit of preparation, we can now specify the thermo-mixed double state $\rho_{\tMD}$ as the unique mixed state defined through the following four properties
\addtolength{\parskip}{-1mm}
\begin{enumerate} 
\addtolength{\parskip}{-3mm}
\item The reduced density matrix obtained by tracing out one side is equal to the thermal density matrix on the other side: ${\rm tr}_L(\rho_{\tMD}) \, = \, \rho_R$.
\item The eigenvalues $j$ of the Casimirs~$J$ are classically correlated between {\small $L$} and  {\small $R$}: the state is an incoherent sum of $J$ eigen states with identical eigenvalues on both sides. 
\item The $s$ quantum numbers that label the different degenerate $J$ eigen sectors are uncorrelated between left and right.
\item The entanglement between the two sides is maximized, subject to properties 1, 2 and~3.
\end{enumerate}
\addtolength{\parskip}{1mm}
These properties uniquely determine the TMD state, and are equivalent to the geometric replica wormhole definition \eqref{tmddef} of the R\'enyi entropies. Property 1) is manifest from the geometric definition \eqref{holotherms} of $\rho_{\tMD}$. Applying the geometric gluing prescription to the geometric representation of the TMD state, it is easy to see that its second R\'enyi entropy ${\rm tr}(\rho_{\tMD}^2)$ equals the  double trumpet partition function \cite{HV:2020}. 
Property 2) follows from the fact that the expectation value of the Casimirs takes the same value everywhere on all the $n$ boundaries of $\Sigma_n$. Properties 1) through 4) combined follow from property 
\addtolength{\parskip}{-1mm}
\begin{enumerate} 
\setcounter{enumi}{4}
\item The $n$-the R\'enyi entropy equals the $n$ wormhole partition function $Z(\Sigma_n)$.
\end{enumerate}
\addtolength{\parskip}{1mm}
The unique state satisfying all these properties is given by an incoherent sum of thermo-field double states
\bea
\label{tmdnew}
\rho_{\tMD}\! \is  \! \sum_{j,s,s'}\; \frac{p_j}{d_j^{\spc 2}} \; 
\smpc |\TFD_j\ra_{ss'}\smpc \la \TFD_j |_{ss'} 
\eea
with $ |\TFD_j\ra_{ss'}$ the thermo-field double state between two $J$ eigen sectors with identical eigenvalues $j$  but with different quantum numbers $s$ and $s'$ on both sides
\bea 
\label{tmdss}
  |\TFD_j\ra_{ss'} \! \is  \sum_m \, \sqrt{q_{jm}\!} \;  | jms\ra_L\,  | jms'\ra_R  
\eea
The formula \eqref{tmdnew} can be rewritten as a dephased incoherent sum of generalized TFD states. Writing out the sums gives 
\bea
\rho_{\tMD} \nspc \is \nspc \sum_{j,m,m',s,s'} \, \frac{p_j}{d^{\, 2}_j}\; \sqrt{q_{jm} q_{jm'}}\; | jms\ra_L\nspc\la jm' s| \otimes | jms'\ra_R\la jm's'|.
\eea
As promised, the $j$ quantum numbers are classically correlated, the $m$ quantum numbers are entangled and the $s$ quantum numbers are uncorrelated between {\small $L$} and  {\small $R$}. 

\subsection{Entropy and mutual information}

\vspace{-.5mm}

We now discuss the entropy and mutual information contained in the above three thermal states. We start with the R\'enyi entropies $S_n = {\rm tr}(\rho^n)$
and then compute the von Neumann entropy via the standard formula $
S_{\rm vN} = - {\rm tr}(\rho\log \rho) \, =\, \lim_{n\to 1} \frac 1 {1-n}\log{\rm tr}(\rho^n).$

The single sided thermal state has R\'enyi entropy
\bea
{\rm tr}(\rho^n_R) \, = \, \sum_j  \, d_j\, \Bigl(\frac{p_j}{d_j}\Bigr)^{\! n}\, \frac{\zzz_j(n \beta)}{\zzz_j(\beta)^n} \nspc \is  \nspc  \frac{Z(n\beta)} {Z(\beta)^n} 
\eea
with $Z(\beta)$ and $Z_j(\beta)$ defined in \eqref{zbetaj} and $p_j$ the probability distribution defined in \eqref{pj}.  From this we find that the von Neumann entropy naturally splits up into three contributions
\bea
\label{srhor}
S(\rho_R) \is  S_p(\beta) + \la S_q(\beta) \ra +  \la \log d_j \ra\\[-11mm] \nonumber
\eea
with\\[-9.8mm]
\bea
S_p(\beta)\nspc \is \nspc - \sum_j \,  p_j \log p_j ,\qquad \qquad\ \ \,  \la \log d_j \ra \nspc = \spc \sum_j \, p_j \log d_j , \\[3mm]
S_q(\beta)\nspc   \is \!- \sum_m q_{jm} \log q_{jm}, \nspc\qquad \qquad \la S_q(\beta)\ra\spc =\spc \sum_j \, p_j\, S_q(\beta)\, .
\eea
The first term $S_p(\beta)$ is the Shannon entropy of the probability distribution among the eigen sectors of the Casimir operators, while $S_q(\beta)$ denotes the entropy within the sector with given $j$ and $s$. The relative size of the three entropy contributions is determined by the relative spectral densities and the range of values of the three quantum numbers $j$, $s$ and $m$.

The $n$-th R\'enyi entropy of the thermo-mixed double state can be read off directly from the explicit expression \eqref{tmdnew} as an incoherent sum of TFD states
\bea
\label{renyimatch}
{\rm tr}(\rho^n_{\tMD}) \nspc \is \nspc  \sum_j  \, d_j{\!}^2\, \Bigl(\frac{p_j}{d^{\, 2}_j}\Bigr)^{\! n}\, \, = \; \sum_j\, \frac {e^{-n\beta E_j} \, \zzz_j(\beta)^n} {d^{\spc n-2}_j\,\spc {Z(\beta)^n\!\!}} \;   
\, = \, \frac{Z(\Sigma_n)}{Z(\beta)^n}
\eea
In the second step, we used the explicit formula \eqref{pj} for the probability distribution $p_j$ and \eqref{ztwo} for the partition function $Z(\Sigma_n)$ of the wormhole geometry. 
From the first expression \eqref{renyimatch} we directly read off the von Neumann entropy 
$S(\rho_{\tMD})\nspc  =\nspc  \spc S_p(\beta) \spc +  2 \spc \la\log d_j\ra$.

It is instructive to compare the von Neumann entropy and mutual information $I_{LR} = S_L + S_R - S_{LR}$
contained in the three types of thermal states.  
Note that the one-sided reduced density matrices obtained by tracing out the other side are in all three cases given by the thermal density matrix \eqref{rhorj}. Hence the sum of the mutual information and total von Neumann entropy for all three states adds up to twice the one-sided thermal entropy.

The factorized thermal state $\rho_L \! \otimes \nspc \rho_R$ has maximal von Neumann entropy, equal to twice  the one-sided thermal entropy, and zero mutual information between the two sides
\bea
S(\rho_L \! \otimes \nspc \rho_R)\! \is\!  2S_p(\beta) \nspc + \la 2S_q(\beta) \ra + \la 2\log d_j \ra\nspc  \qquad \qquad\ I_{LR}(\rho_L\! \otimes\nspc \rho_R)\smpc = \smpc 0.
\eea
The thermo-field double, on the other hand, has zero von Neumann entropy, but maximal mutual information equal to twice the one-sided thermal entropy
\bea
\quad S(\rho_{\tFD}) \nspc \is\spc  0, \qquad \qquad \ \  I_{LR} (\rho_{\tFD})\spc = \spc 2S_p(\beta) \nspc + \la 2 S_q(\beta) \ra + \la\nspc 2 \log d_j \nspc \ra\nspc. 
\eea
By contrast, the entropy and mutual information of the TMD state splits up~as
\bea
\label{stmd}
S(\rho_{\tMD}) \nspc \is\nspc   S_p(\beta)+ \spc \la 2 \log d_j\ra \spc , \qquad \qquad\quad\ I_{LR}(\rho_{\tMD}) \spc = \spc  S_p(\beta) +  \spc \la 2S_q(\beta)\ra\spc .
\eea
We see that the $S_p(\beta)$ contribution is split equally between the entropy and mutual information.  This reflects that the TMD is an incoherent sum of terms with equal Casimir values on both sides. The $s$ quantum numbers, on the other hand, are uncorrelated and do not contribute to the mutual information, while the $m$ quantum numbers are fully entangled and do not contribute to the entropy. So depending on which of the three terms dominates, the result for $S(\rho_{\tMD})$ or $I_{LR}(\rho_{\tMD})$ is closer to zero, once or twice the single sided entropy \eqref{srhor}.

\bigskip

\section{Conclusion}
\vspace{-2mm}

We have introduced a geometric path integral definition for wormhole partition functions in a wide class of quantum systems. The basic idea is that for many interesting quantum systems, the symplectic two-form on the quantizable orbits is not exact and that the path-integral formula for such systems involves integrating over a two-dimensional surface. 
While the construction in this paper may seem special and unrelated to holography, it applies to many known systems including any 2D CFT with Virasoro symmetry. As outlined in Appendix B, the phase space of 2D Virasoro CFT is given by the Teichm\"uller space of constant curvature metrics and applying the general reasoning used in section 2 in essence amounts to a derivation of AdS${}_3$ gravity from the CFT \cite{HV:1989}.  The 2D CFT application gives a useful illustration of our set up and of the role of the three quantum numbers $j$, $m$ and $s$.

We have shown that partition function $Z(\Sigma_n)$ of the $n$-fold wormhole geometry does not factorize and is equal to the $n$-th R\'enyi entropy of a specific thermal mixed state $\rho_{\tMD}$.   The thermal mixed double state incorporates three sectors with different types of quantum statistical behavior: classically correlated quantum numbers $j$, entangled quantum numbers $m$ and classically uncorrelated quantum numbers $s$. 
Among the three, only the entangled quantum numbers $m$ have a partition function that factorizes into a product. The partition function of the classically (un)correlated quantum numbers $j$ and $s$ do not factorize. Our reasoning  thus  illuminates the quantum statistical implications of the non-factorized wormhole contributions.

The observation that replica wormholes have a natural place in ordinary quantum systems sharpens the question about what truly distinguishes holographic theories from non-holographic theories.  Holographic quantum systems typically have a chaotic spectrum without exact degeneracies, but they may have a large number of approximately degenerate energy states. As prime example of a candidate holographic quantum system is the SYK model \cite{Kitaev,MaldacenaStanford:2016}.  By exploiting the $SO(N)$ symmetry of the non-interacting or the disorder averaged SYK model, our methods could in principle be applied to introduce a microscopic notion of SYK wormhole configurations and define a microscopic notion of a TMD state. We leave this application for future study.

\bigskip

\section*{Acknowledgements}
\vspace{-2mm}

It is a pleasure to thank Ahmed Almheiri, Akash Goel, Luca Iliesiu, Juan Maldacena, Thomas Mertens, Douglas Stanford, Joaquin Turiaci, Mengyang Zhang and Wenli Zhao for helpful discussions and comments. This research is supported by NSF grant number PHY-1914860.

\bigskip

\def\lbl{{\raisebox{-.5pt}{\large$($}}}
\def\rbl{{\raisebox{-.5pt}{\large$)$}}}

\appendix
\section{Wormholes in QM: Three examples}

\subsection{Coupled oscillator}

\vspace{-1mm}

Consider the quantum system of a coupled harmonic oscillator, described by the four-dimensional phase space $(p_i , q_i)$  with $i=1,2$, with Poisson brackets $\{p_i,q_j\} = \delta_{ij}$ and Hamiltonian  $
H  = \half \spc p_{i}^2 + \half \spc k_{ij} q^i q^j.$
This simple system does not have a holographic gravity dual, but as we will see, we can still define a wormhole partition function. 

To make the coupled oscillator look holographic, we will choose to parametrize the phase space in terms of a new set of variables, as follows. Let $(a^*_\alpha,a_\alpha)$ denote the raising and lowering operators of the two normal modes with Poisson bracket $\{ a^*_\alpha,a_\beta\} = \delta_{\alpha\beta}$. The Hamiltonian can then be cast in the form
\bea
H \is    \spc \omega \spc {J}  + \, \epsilon\spc {X}_3,
\eea
with  ${J} \spc = \spc  {a}^*_\alpha {a}_\alpha$ and ${X}_a = \half \, {a}^*_\alpha \sigma_a^{\alpha\beta}\spc {a}_\beta, $, 
with $\sigma_a$ the Pauli matrices and $\omega$ and $\epsilon$ some constant frequencies. $J$ and $X_a$ satisfy\footnote{Here we raise and lower indices with the 3-d Kronecker symbol $\delta_{ab}$}
\bea
\label{jxdef}
J^2 \! & \! \equiv \! &\! X^a X_a.
\eea 
We now choose phase space coordinates $(\tau, X^a)$, where $\tau$ is the  canonical conjugate variable to $J$. The Poisson brackets are
\bea
\label{poissonb}
\{J,\tau\} = 1, \qquad \{ X^a,X^b\} =  \epsilon^{ab}{\!}_c \; X^{c},\qquad 
\{J,X^a\}  
\, = \, 0\, .
\eea
 The $X^a$  variables are $SU(2)$ symmetry generators and $J^2$ is the $SU(2)$  Casimir. Note that $J$ is indeed a conserved quantity and  $(J,\tau)$ are action angle variables. In the analogy with holographic CFT, the $X^a$ play the role of the generators of the conformal group, $J$ is the Casimir operator that measures the conformal dimension of the primary field of a conformal family, and $H$ is the total conformal scale dimension. 
 
We can encapsulate the Poisson bracket algebra \eqref{poissonb} by means of a symplectic form $\Omega$, given by the  following closed 2-form on the phase space $(X,\tau)$ 
\bea
\Omega\spc = \spc
\frac 1 2 \, \omega_{ab} \spc dX^a\!\!\smpc \wedge \!\smpc d X^b   \spc +\spc dJ\!\smpc\wedge\!\smpc d\tau, 
\eea
where $\omega_{ab}$ is the inverse of the matrix $\omega^{\spc ab} = \epsilon^{abc}X_c$, and with $J$ as defined in \eqref{jxdef}. The allowed integer spectrum of $J$ follows from the geometric quantization condition that  the integral  of the symplectic form $\omega$ over any 2-cycle must be $2\pi$ times an integer. 

We can restrict the phase space to a single quantizable orbit by performing a symplectic reduction: we set $J =j$ with $2j\in \mathbb{N}$ and mod out by the motion $\tau \to \tau + c$ generated by $J$. The resulting  orbit takes the form of a two sphere $X^a X_a = j^2$ with radius $j$. Upon quantization, it gives rise to an $SU(2)$ representation of spin $j$.
Hence, the quantum theory obtained by promoting the above Poisson brackets to commutators has an energy~spectrum 
\bea
E_{jm} \is  \gamma j + \epsilon \spc m, \qquad 2j \in \mathbb{N} , \qquad -j\leq m \leq j 
\eea
that indeed matches with the spectrum of two harmonic modes with frequency $\omega_\pm = \half(\gamma \pm \epsilon)$. 

After performing the reduction procedure, the symplectic form simplifies to 
\bea
\label{reducedomega}
\omega  = \frac 1 2 \, \omega_{ab} \spc dX^a\!\!\smpc \wedge \!\smpc d X^b .
\eea 
This reduced symplectic form is still closed but no longer exact $\omega \neq d \alpha$. Hence the integral $\int_\Sigma \omega$ over an open 2D phase space region $\Sigma$ with boundary $\partial \Sigma$ can not be written as an integral $\int_{\partial \Sigma} \alpha$ over the boundary. 

We now define the wormhole partition function of this toy model.  Let $\Sigma$ denote a 2D surface with boundary $\partial \Sigma$ parametrized by a (euclidean) time coordinate $t$. 
We associate a partition function $Z(\Sigma)$ to $\Sigma$ via the following path integral formula
\bea
\label{zsigmat}
 Z(\Sigma)\is  \int \! \spc [dX d\tau\spc] \,\spc e^{\mbox{\footnotesize $ \frac{i}{\hbar} \smpc S_{\Sigma}[X,\tau\spc ]$}}\\[4.5mm]
S_{\Sigma}[X,\tau]\, =\, \int_\Sigma\! \Omega \!\! & -&\!\!\!\! \oint_{\partial \Sigma} \!\!  H dt \spc = \spc \int_\Sigma \nspc \omega  
 \spc   
+\spc \oint_{\partial \Sigma} \!  ( J d\tau \smpc -\smpc H dt\smpc ) \, 
 \eea
with $J$ defined in \eqref{jxdef}. Here $X$ and $\tau$ are 2D fields defined on the surface $\Sigma$. The boundary values of $\tau$ are integrated over, while for $X$ we will consider either Neumann or Dirichlet boundary conditions at $\partial \Sigma$. Since the bulk action is an integral of a closed two-form, it defines a unique functional of the boundary values of $X$, up to integer multiples of  $2\pi i \hbar$. This property ensures that the functional integral \eqref{zsigmat} is uniquely defined. 

The Hamiltonian $H$ does not depend on the angle variable $\tau$. The dual variable $J$ is therefore a conserved quantity: $\tau$ acts as a  lagrange multiplier that imposes the conservation condition $dJ=0$ along each boundary component of $\Sigma$. The $\tau$ variable satisfies the periodicity condition that $ \oint J d\tau$ is an integer multiple of $2\pi i \hbar$ for all quantizable orbits.

\subsection{Particle on a group II}
\vspace{-1mm}

We summarize the computation of the $Z(\Sigma_n)$ for a particle on a group \cite{Blau:1993} \cite{Iliesiu:2019b}. First we turn off the Hamiltonian. 
The Hilbert space is isomorphic to the space spanned by all functions on a group $G$. This space decomposes into irreducible representations as a direct sum $\oplus_{j} d_j R_j$ with $d_j={\rm dim} R_j.$ The partition function twisted by a unitary operator $U(g)$ then decomposes into a sum of characters $\chi_j(g) = {\rm tr}_{R_j}(U_{R_j}(g))$ via
\bea
\label{zg}
Z(g)\nspc  \is\nspc {\rm tr}(U(g)) \spc = \spc \sum_j\, d_j \, \chi_j(g) \, 
\eea
We can represent this as a path-integral \eqref{zsigmanew} defined on a disk $D$ with $e^{-\oint dt H}$ replaced by $U(g)$. Equation \eqref{zg} has a simple significance. By virtue of the identity
\bea
\label{deltag}
\delta(g-1) \is \sum_j\, d_j \, \chi_j(g)
\eea
we see that $Z(g) = \delta(g-1)$. This is as expected: the theory with $H=0$ is topological and the boundary of $D$ is contractible. So a twisted boundary condition implies the presence of a defect. In the absence of this defect, the partition function $Z(g)$ with $g\neq 1$  has to vanish.
 
Now let us turn on the Hamiltonian given in \eqref{gactt}. When acting on a given representation $R_j$, we have $e^{-\oint dt H} = e^{-\beta E_j} U_{R_j}(g_\beta)$ with $g_\beta = e^{-\beta A^a \tau_a}$.  The finite temperature partition function thus decomposes into a sum of group characters as follows
\bea
\label{zgbetat}
Z(\beta)\nspc \is\nspc \sum_j \, d_j\,  \chi_j(g_\beta) \, e^{-\beta E_j}
\eea
Note that this partition function is always non-zero. We would like to do the same computation for the partition function associated with the $n$-fold replica wormhole geometry~$\Sigma_n$. The calculation is quite standard and makes use of the following two identities
\bea
\label{idone}
\int\! dg \, \chi_{j_1} (g) \chi_{j_2} (h g^{-1}) \is \left\{ \begin{array}{c} { \delta_{j_1,j_2} \, \raisebox{-1pt}{\Large $\frac{1}{d_j}$} \, \chi_{j_1}(h)\ \ }\\[2mm] { \delta_{j_1,j_2}\quad \ \raisebox{-1pt}{if\ \  $h=1$}} \end{array}\right. \\[2mm]
\int\! dh \, \chi_{j}(hg_1 h^{-1} g_2)\is 
\, 
\frac{1}{d_j}\,\chi_j(g_1) \chi_j(g_2)
\label{idtwo}
\eea

As a next step, consider the following partition function
\bea
Z(g,\beta) \is \sum_j  \,
\chi_j(g)\spc \chi_j(g_\beta)  \, e^{-\beta E_j}
\eea
As we will see shortly, it represents the partition function on an annulus with finite temperature boundary conditions on one side and an insertion of $U(g)$ on the other side.

Using the first identity \eqref{idone}, we can write $Z(\beta)$ in \eqref{zgbetat} as a convolution of the mixed annulus partition function and the twisted disk partition function \eqref{zg}. We interpret this as a geometric gluing formula
\bea
\label{glueone}
Z(\beta) \nspc \is \nspc \int\! dg \, Z(g) \, Z(g^{-1}, \beta)\qquad \qquad
\raisebox{-5mm}{$ \includegraphics[scale=.42]{bhdisk.pdf}$}g \ \ g^{\! -1}\!\raisebox{-6mm}{$ \includegraphics[width=.9cm, height=1.4cm]{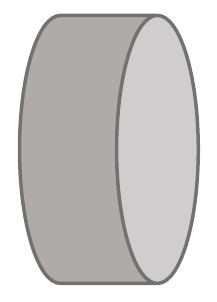}$} \beta\quad \mbox{\normalsize $=$} \quad \raisebox{-5mm}{$ \includegraphics[width=1.27cm, height=1.3cm]{bhdisk.pdf}$} \beta
\eea
The integral over $g$ represents a sum over a complete set of intermediate states.

We can generalize the gluing formula \eqref{glueone} and write the finite temperature partition function $Z(\Sigma_n) = Z(\beta,\beta,..,\beta)$ on the $n$-fold trumpet  $\Sigma_n$  as an $n$-fold convolution of the zero temperature partition function $Z(g_1,g_2,..,g_n)$  on $\Sigma_n$ with $n$ annulus partition functions
\bea
\label{zebeta}
Z(\beta,\beta,..,\beta)\, =  \, 
\int\! dg_1...dg_n \, Z(g_1,g_2,...,g_n)\,  Z(g_1^{-1}\!,\beta)\, ...\, Z(g_n^{-1}\!,\beta) \, .
\eea
Through our general formula \eqref{zsigmanew}, $Z(g_1,g_2,...,g_n)$ defines a partition function of a BF gauge theory. This in turn we can equate with the volume
\bea
Z(g_1, g_2, ...,g_n) \is {\rm Vol}(g_1,g_2,...,g_n)
\eea
 of all flat $G$ bundles with given holonomy around the closed one-cycles surrounding each boundary component. This volume is computed as follows.

Each holonomy is given by the trace of a Wilson line, and thus only specifies the conjugacy class $[g_k]$ of the corresponding group element~$g_k$.  In other words,  the path ordered exponential of the BF gauge field around the boundary circles are of the form $h_k g_k h_k^{-1}$ for some set of group elements $h_k$. 
Similar as discussed above for the disk, the loop surrounding all boundary components is contractible. Therefore the product of the path-ordered exponentials must be trivial. This specifies the space of flat $G$ bundles on $\Sigma_n$ with given holonomies $[g_k]$ as the space of all group elements $h_1, h_2, ..., h_n$ such that 
\bea
\label{constraint}
h_1 g_1 h_1^{-1} \! \cdot h_2 g_2 h_2^{-1} \cdot \cdot \cdot h_n g_n h_n^{-1} \is 1.
\eea
The volume of this space can be directly evaluated, using the formulas \eqref{deltag} and \eqref{idtwo}
\bea
& & {\rm Vol}({g_1,g_2,...,g_n})= \int \! \spc dh_1\, dh_2 ... dh_n \; \delta(h_1 g_1 h_1^{-1} \! \cdot h_2 g_2 h_2^{-1} \! \cdot \cdot \cdot h_n g_n h_n^{-1}\!\! - 1)  =  \nonumber \\[-1.5mm]\\[-1.5mm] \nonumber
& & \sum_j  d_j \, \int \!\!\spc dh_1\spc ... dh_n \spc \chi_j(h_1 g_1 h_1^{-1} \! \cdot \cdot \cdot h_n g_n h_n^{-1} )\spc = \, \sum_j \, \frac 1 {d_j^{\spc n-2}\!}\; \, \chi_j(g_1) \spc \chi_j(g_2) \, ...\, \chi_j(g_n)
\eea
Applying the $n$-fold gluing formula \eqref{zebeta} and using the identity \eqref{idone} gives
\bea
\label{zchecktt}
Z(\Sigma_n) \nspc \is \nspc \sum_j \, \frac {e^{-n\beta E_j}\spc \chi_j(g_\beta)^n} {d_j^{n-2}\!}
\eea
This matches with the announced result \eqref{zpredict} for the partition function for the $n$-fold replica wormhole geometry via the identification of $\chi_j(g_\beta)$ with the thermal partition function $\zzz_j(\beta)$ for given values of the Casimir operators $J_I$. 

The above result can also be derived by using the isomorphism between the Hilbert space  ${\cal H}$ = Fun($G$)  and the group algebra ${\bf C}[G] = \{x=\sum_{\gamma} x(\gamma)\cdot \gamma\}$ with  $x(\gamma) \in {\bf C}$. This isomorphism induces a multiplication rule on Fun$($G$)$ via $x \cdot y = \sum_{\gamma_1, \gamma_2} x(\gamma_1) y(\gamma_2) \gamma_1 \gamma_2$. Hence we can think of the Hilbert space as as a direct sum of $d_j \times d_j$ matrices
\bea
\label{decom} {\cal H}  \! & \! \cong \! & \! \bigoplus_{j} {\rm
Mat}( d_j)\, .
\eea
To derive formula \eqref{zchecktt} we represent the trace over the Hilbert space ${\cal H}$ by a matrix integral over elements $X$ of the group algebra  
 \bea
\bigl\langle Z(\beta)^n \bigr\rangle\! \is\! \int\! dX\, Z(\beta)^n\qquad \quad Z(\beta) =
 \tr\lbl e^{-\beta H_0(X)}U(g_\beta)\rbl
 \eea
We the evaluate this integral perturbatively, while keeping only the leading connected contribution due to planar Feynman diagrams with the topology of $\Sigma_n$. The result \eqref{zchecktt} then follows from the usual 't Hooft counting.

\def\loplus{\raisebox{1pt}{\Large $\oplus$}}

\subsection{Schwarzian Quantum Mechanics}
\vspace{-1mm}

It is natural to consider the application of our set up to Schwarzian quantum mechanics. 
 Generalizing our construction to this case amounts to a derivation of JT gravity from the Schwarzian.  A new element is that the spectrum is continuous: the thermal partition function of the Schwarzian is given by an integral over the spin $j= -\frac 1 2 + i \lambda$ of $SL(2,\mathbb{R})$ representations
\bea
\label{zbetat}
Z(\beta) \spc =\spc {\rm tr}(e^{-\beta H}) \nspc \is \!  e^{S_0} 
\int\! d\mu(\lambda)\, e^{-\beta \lambda^2} , \qquad \qquad d\mu(\lambda) = \lambda \sinh 2\pi \lambda
\eea
The above formula should be compared with equation \eqref{zgbeta} for the partition function of a particle on the group manifold.
The spectral density in \eqref{zbetat} is the appropriate Plancherel measure on the space of representations \cite{Iliesiu:2019}.  
The $SL(2,\mathbb{R})$ symmetry is a gauge symmetry in Schwarzian quantum mechanics, so compared to \eqref{zgbeta} we must set $\chi_j(g_\beta) = 1$. 
The constant ground state entropy $S_0$ is added by hand and corresponds to the overall constant zero mode of the dilaton in JT gravity. This indicates that we should treat the factor of $e^{S_0}$ as indicating the presence of an exact degeneracy. It indeed behaves in the same way as the degeneracy $d_j$ of the representations in the previous sections.

\begin{figure}[t]
\begin{center}
\vspace{-1mm}
\includegraphics[scale=.3]{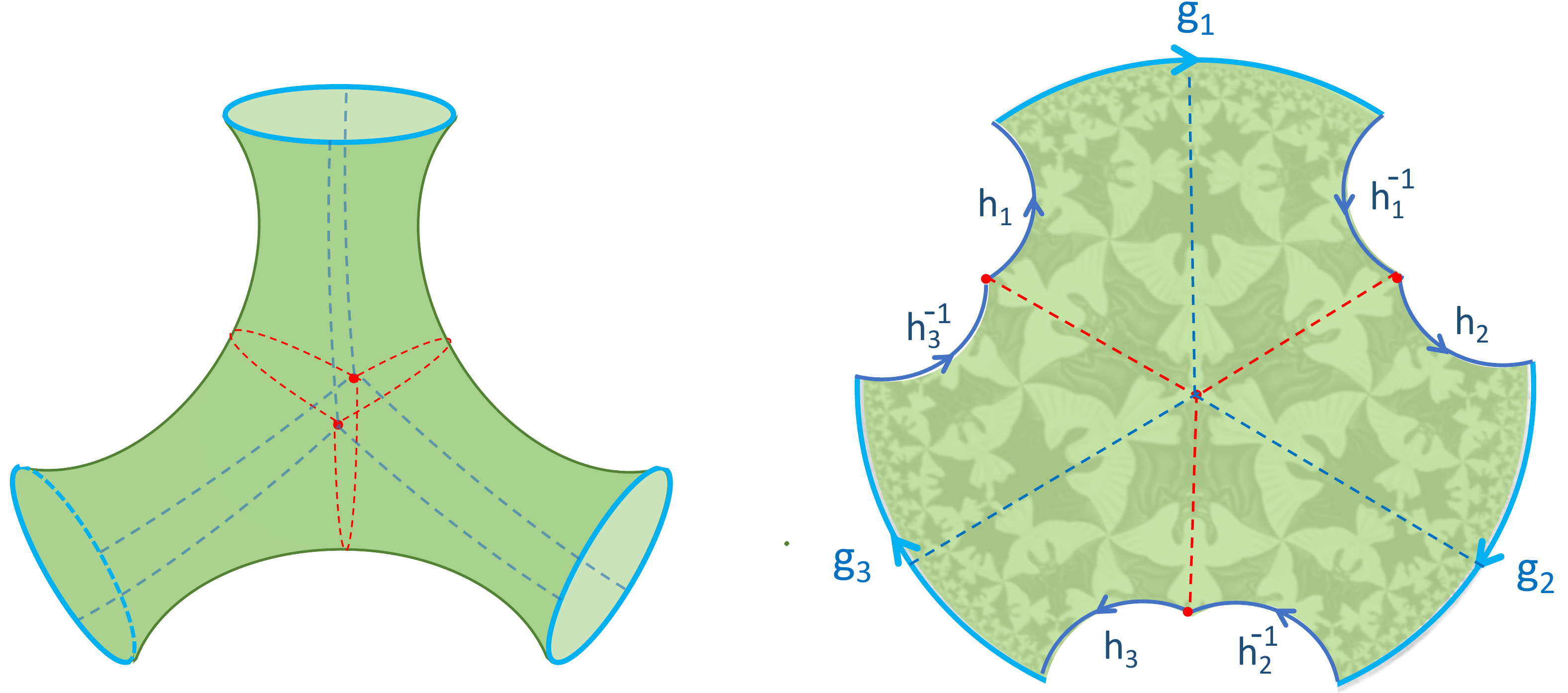}
\caption{\small Replica wormhole geometry for the third R\'enyi entropy represented as a quotient of the Poincar\'e disc. The side edges are all labeled by $SL(2,\mathbb{R})$ group elements.}
\end{center}
\vspace{-5mm}
\end{figure}

The JT gravity partition function is given by the volume of the moduli space of constant curvature metrics on the 2D manifold.  The $n$ fold wormhole geometry $\Sigma_n$ has $-3\chi_{\Sigma_n} = 3n-6$ shape moduli.  We can represent $\Sigma_n$  as a quotient of the Poincar\'e disc, as indicated for $n=3$ in figure 2. Each edge segment follows a geodesic connecting two end points related via an $SL(2,\mathbb{R})$ M\"obius transformation. As before, the group elements must satisfy the holonomy constraint. 
  In our context, $\Sigma_n$ comes with a natural marking, that indicates the way it was glued together from elementary geometries.  In particular, each boundary circle  comes with two marked points that indicate how it was obtained via gluing together two half-circles. The markings break the invariance under the mapping class group of $\Sigma_n$. With this consideration, we should {\it not} divide out the modular group and integrate over the full Teichm\"uller space.

The calculation now follows the same steps as in section 4.2, yielding the following result for the $n$-fold wormhole partition function
\bea
\label{zcheckt}
Z(\Sigma_n) \nspc \is \nspc e^{-(n-2) S_0} \int d\mu(\lambda)  \, e^{-n\beta E_\lambda}\spc = e^{-(n-1)S_0} Z(n\beta) 
\eea
This expression is different from the one derived by from the Mizakhani volume of the moduli space of constant curvature metrics on $\Sigma_n$ and from the connected contribution from the corresponding double scaled matrix model. As explained above, the difference is that in our case, by virtue of our replica Ansatz, the trumpet geometry has a natural marking and we are therefore instructed to integrate over the full Teichm\"uller space. The formula \eqref{zcheckt} can be written as the $n$-th R\'enyi entropy of the following thermo-mixed double state
\bea
\rho^{(0)}\is\! e^{-2S_0} \sum_{s,s'} |s\rangle_L \langle s| \otimes |s'\rangle_R\langle s'| \nonumber\\[-4mm]
\rho_{\tMD} \, = \, \rho^{(0)} \otimes \rho^{(1)}_{\tMD}\qquad \qquad\qquad && \\[-1mm] \nonumber 
\rho^{(1)}_{\tMD}\!\! \is\! \int\!\nspc d\mu(\lambda) \spc e^{-\beta E_\lambda}  |\lambda\rangle_L \langle \lambda | \otimes |\lambda\rangle_R\langle \lambda| 
\eea
As before, we see that the quantum number $\lambda$ that labels the representations is classically correlated, whereas the $s$ label is uncorrelated between {\small $L$} and {\small $R$}. In this case, there is no analog of the $m$ quantum number, since the $SL(2,\mathbb{R})$ symmetry is gauged.

 \section{Wormholes in  2D CFT}
\vspace{-2mm}

In this Appendix we explain how the story in the main text relates to holography and the AdS/CFT correspondence. We will show how the same reasoning applied to an arbitrary 2D conformal field theory with Virasoro symmetry results in a geometrical path integral formulation in terms of 2+1-dimensional AdS gravity. In this sense, our reasoning can be viewed as a derivation and premonition of AdS${}_3$/CFT${}_2$: the content of this section was known and described in more detail  in \cite{HV:1989, Verlinde:1989}. Our discussion below illustrates how the preceding arguments are intertwined with the origin of the holographic correspondence.

\subsection{Geometric phase space of 2D CFT}
\vspace{-1.5mm}

A 2D CFTs is invariant under the group of conformal transformations $
(z,\bar{z}) \, \to \, (f(z),\bar{f}(\bar{z}))$
 given by the product 
 of the left times right Virasoro symmetry group. A more common name for the Virasoro group is Diff($S^1$), the group of diffeomorphisms of the circle.
To simplify our notation, we will temporarily focus on one chiral half of the full conformal group. We will return to the full non-chiral story later. 

The group ${{\rm Diff} (S^{\smpc 1})}$ is a Poisson manifold with a natural symplectic form, designed such that the space of functions satisfy Poisson brackets identical to the Virasoro algebra  \bea
[L(\xi_1),L(\xi_2)] = L([{\xi}_1,{\xi}_2])+ \frac{c}{24\pi }\oint {\xi}_1 {\xi}^{'''}_2
\eea
with $[\xi_1,\xi_2] = \xi_1\xi'_2 - \xi_2 \xi'_1$ the usual Lie bracket of two vector fields. Quantization of all quantizable symplectic orbits of ${{\rm Diff} (S^{\smpc 1})}$ produces a Hilbert space that contains all unitary Virasoro modules. The CFT Hamiltonian is given by $L_0$, and the energy spectrum has a structure that looks similar to the one considered in the previous sections. The Hilbert space is a direct sum of Virasoro modules, labeled by primary states $|\Delta\ra$, spanned by all descendent states 
\bea
|\Delta,n \ra\, =\, \prod_i L_{-n_i} |\Delta\ra, \qquad
\mbox{with energy} \qquad
E_{\Delta,n} \is \Delta + \sum_{i} n_i.
\eea
This spectrum should be compared with equation \eqref{esplit}. For a  holographic CFT in the Cardy regime, the spectrum of primary operators is very dense compared with the integer energy gaps between the successive descendent states.

The Casimir of ${{\rm Diff} (S^{\smpc 1})}$ takes a constant value for each Virasoro module. The restricted phase space with fixed value of the Casimir are called co-adjoint orbits. A useful characterization of these orbits is in terms of the Teichm\"uller space of the disk~$D$. The Teichm\"uller space of a surface can be characterized in two equivalent ways. First, we can define ${\rm Teich}(\Sigma)$ as the space of constant curvature metrics, modulo the subgroup of diffeomorphism  continuously connected to the identity. Locally, we can write a constant curvature metric as $
ds^2 = \frac{df d\bar{f}}{(1-|f|^2)^2}$
with $f(z,\bar{z})$ some arbitrary function of the local complex coordinates. Equivalently, we can define the Teichm\"uller space as the universal covering space of the moduli space of complex structures on a surface. We will make use of both definitions below.

Consider an annular region $A \! =\! \{ z; a\!\leq\! |z|\! \leq\! b\}$ to which we attach a disk $D\! = \! \{ w ; |w|^2 \! < \!1\} $ via the identification $(w,\bar{w})\! =\! (f(z),\bar{f}(\bar{z}))$. Two maps $f$ and $\tilde{f}$ are equivalent if $\tilde{f} \circ f^{-1}$ extends to a holomorphic function on $D$. Teich($D$) is defined as the set of equivalence classes of the maps $(f,\bar{f})$. 
We can define a transitive action of ${{\rm Diff} (S^{\smpc 1})}$ on Teich($D$) via $(f,\bar{f}) \to (F\circ f, \bar F \circ \bar f)$, where $(F,\bar{F})$ are restricted such that the circle $|f|^2 =1$ is mapped to itself. By definition, the $SL(2,\mathbb{R})$ subgroup of global conformal transformation on $D$ acts trivially on Teich$(D)$. 
We conclude that the Teich\"uller space of the disk 
is equal to the special co-adjoint orbit of the identity representation ${\rm Teich}( D) = {{\rm Diff} (S^1)}\smpc / \smpc {SL}(2,\mathbb{R})$.
This identification underlies the relation between the vacuum state $|0\ra$ and the CFT path integral on the disk. The co-adjoint orbits for other Virasoro representations with $\Delta>0$ take the form of the Teichm\"uller space 
${\rm Teich}( D_*)= {{\rm Diff} (S^1)}\smpc / \smpc {S^1}$ of the punctured disk $D_*$ with a curvature singularity at the origin.
The symplectic form on these orbits is designed so that the resulting Poisson brackets are identical to the Virasoro algebra with central charge~$c$. The natural symplectic form of Teichm\"uller space is known as the Weil-Peterson form. This WP form $\Omega_{\mbox{\tiny WP}}$ on  ${\rm Teich}( D_*)$ equals the symplectic form on the associated co-adjoint orbit.

 To write  $\Omega_{\mbox{\tiny WP}}$, we can use the characterization of Teichm\"uller space as the space of constant curvature metrics. It will be useful to adopt a first order formulation and parametrize the 2D geometry  by two frame fields $e^\pm$ and a spin connection $\omega$. Teichm\"uller space then equals the space of solutions to the constant curvature equations
\bea
\label{flatc}
d\omega\spc + \spc \spc e^+ \!\nspc \wedge\nspc e^- \, = \, 0, 
\qquad \qquad  de^\pm \mp\omega \nspc \wedge \nspc e^\pm \! \is 0,
\eea
modulo diffeomorphisms and local Lorentz transformations, or equivalently,  the space of flat $SL(2,\mathbb{R})$ gauge fields $A = (e^+,e^-,\omega)$ modulo gauge transformations. 

For the case of the punctured disk $D_*$, the flatness conditions \eqref{flatc} are valid everywhere except at the curvature singularity at the center of the disk. Hence the trace of the holonomy 
\bea
C(A) = {\rm tr} \bigl( {\rm P} \exp \oint_* A \bigr) 
\eea
around the puncture can be non-trivial. For the Cardy regime, the holonomy is an element of an hyperbolic conjugacy class;  geometrically, the puncture becomes a flaring end of a trumpet. The holonomy $C(A)$ measures the minimal geodesic length $\ell$ of a loop surrounding the trumpet and specifies the conformal weight $\Delta$ of the corresponding co-adjoint orbit via
\bea
\label{cdelta}
C(A) = 2\cosh\Bigl(\frac{\ell_\Delta}{2} \Bigr) \equiv C_\Delta \qquad \qquad \frac{\ell_\Delta}{2\pi} = \sqrt{\frac{24\Delta}{c}}.
\eea
In the ${{\rm Diff} (S^{\smpc 1})}$  description, the holonomy is introduced by allowing the diffeomorphism $F$ to return to itself up to a M\"obius transformation $F \to \frac{a F+b}{cF+d}$ after going around the~$S^1$.

From here on, let $D_{\Delta}$ denote the disk with a curvature singularity at the center with holonomy $C(A) = C_\Delta$.
 In terms of the first order formulation, the Weil-Peterson symplectic form on Teich($D_\Delta$) is now given by the following simple expression
\bea
\label{wp}
\Omega_{\mbox{\tiny WP}} \is \frac{k}{4\pi} \int_{D_{\! \Delta}} \! (\delta e^+ \!\nspc \wedge\nspc \delta e^- + \, \half \, \delta \omega\nspc \wedge \nspc \delta \omega) \qquad \quad k = \frac{c}{6}.
\eea
Here $(\delta e^+,\delta e^-,\delta \omega)$  denote one-forms on the space of flat gauge fields $(e^+,e^-,\omega)$ satisfying the holonomy condition \eqref{cdelta}, modulo gauge transformations. The WP two-form \eqref{wp} on Teich($D_\Delta$) coincides with the symplectic form on the co-adjoint orbit of Diff($S^1$). Quantization of this space produces a Virasoro algebra with central charge $c$ and conformal weight~$\Delta$.

The full CFT has both a left and right moving Virasoro symmetry. So the complete phase space of the CFT consists of the union over the full CFT spectrum of left and right conformal dimensions ($\Delta,\bar\Delta$) of factorized co-adjoint orbits of the form 
\bea
\label{phasespace}
{\rm phase \ space \ of \ CFT }\,  \is\, \raisebox{-10pt}{\LARGE${\cup}\atop{\raisebox{8pt}{\scriptsize $\Delta,\bar{\Delta}$}}$}\; {\rm Teich}(D_{\nspc \Delta}) \times {{\rm Teich}({D}_{\nspc \bar\Delta})}.
\eea
Each ${\rm Teich}(D_{\nspc \Delta})$  has its own WP symplectic form, given in 
$SL(2,\mathbb{R})$ covariant notation~by
\bea
\label{wpt}
 \Omega_{\mbox{\tiny WP}} \is \frac{k}{4\pi} \int_{\strut_{{}^{\mbox{\scriptsize $D_\Delta$}}}} \hspace{-2mm} {\rm tr}( \delta A \wedge \delta A) \qquad \quad \left\{\begin{array}{c}  {F(A)\spc =\spc  \, 0}\\[2mm]
  {\ \, C(A) \spc = \spc C_\Delta}\end{array}\right. 
\eea
 By design, quantizing this full phase space produces the complete spectrum of the CFT.\footnote{\addtolength{\baselineskip}{.5mm} The Virasoro algebra can be derived from the quantization of  Teich($D_{\nspc \Delta})$ as follows \cite{HV:1989, Verlinde:1989}.  First one promotes the Poisson bracket derived from  the Weil-Peterson form \eqref{wpt} to a commutator algebra 
$[A_i^a(x_1), A_j^b(x_2)]=  \frac{4\pi}{k} \, \epsilon_{ij} \delta^{ab} \delta(x_{12})$. The flatness condition $F(A) = 0$ is imposed as a constraint on the wave functionals. The key next step is to let the wave functionals depend on $e_+$ and one of the chiral components $\omega_z$ of $\omega$ \cite{HV:1989}. The other components act as functional derivatives via $e^-_\alpha =\spc \frac{4\pi}k \spc {\epsilon_{\alpha\beta}}\frac{\delta\ \,}{\delta e{}^+_\beta\!}$ and $\omega_{\bar{z}}  =\spc \frac{4\pi}k \spc \frac{\delta\ \,}{\delta \omega_z\!}$ . The three flatness equations \eqref{flatc} then become three functional differential equations, which can be reduced to a single linear differential equation equivalent to the Virasoro Ward identity. The wave functionals are identified with the generating function of correlation functions of the stress tensor. }

The above description of the phase space of the 2D CFT follows the same pattern as in the previous sections. The role of the $X^a$ operators is now taken over by the Virasoro generators $L_n$, while the holonomy operator $C(A)$ is a central function of the Poisson algebra and thus indeed represents a Casimir operator of the Virasoro algebra. Moreover, we can define an analog of the Wilson operators $W(\tau)$ that change the eigen value of the Casimir, by considering the open Wilson line operators $W(A) = {\rm P}\exp\int^*\! A$ that start at the boundary of the disk and end at the puncture at the center of the disk. Using the explicit expression \eqref{wp} of the WP form, one can show that such open Wilson lines indeed create or change the curvature singularity at the puncture, and thus modify the holonomy $C(A)$ according to an appropriate fusion rule. In the context of rational CFTs, the operators $C(A)$ and open Wilson lines $W(A)$ are known as Verlinde operators that generate the fusion algebra.

\subsection{Wormhole partition function in 2D CFT}
\vspace{-1mm}

With this preparation, we can now apply the  wormhole discussion developed for quantum mechanics in section 2 to 2D CFTs. 
Consider two 2D CFTs on a circle with the same spectrum and same classical phase space \eqref{phasespace}-\eqref{wpt}. We would like to write the wormhole partition function $Z_{\rm CFT}(\Sigma_n)$ defined via the path integral prescription \eqref{zsigma}. Via the same reasoning as in section 2, we learn that this partition function $Z_{\rm CFT}(\Sigma_n)$ takes the form of a sum over the spectrum of Casimir eigen values, labeling the left- and right primary weights 
\bea
\label{zcft}
Z_{\rm CFT}(\Sigma_n) \is \sum_{\Delta,\bar{\Delta}}  
Z_\Delta(\Sigma_n) \bar{Z}_{\bar{\Delta}}(\Sigma_n),
\eea
where the chiral partition functions $Z_\Delta(\Sigma_n)$ with fixed primary weight are given by
\bea
\label{zdelta}
 Z_\Delta(\Sigma_n) \is \int_{\strut{\raisebox{-6pt}{\scriptsize ${F(A)=0\ } \atop {C(A) = C_\Delta}$}}}\hspace{-9mm}\! [dA] \ e^{- \int_{\Sigma_n} \Omega_{\mbox{\tiny WP}}\,} 
\eea
Here we temporarily turned off the boundary Hamiltonian $h(X)$; we will reinstate it later.

We would like to compare the above path-integral formula with the holographic replica wormhole prescription in 2+1 AdS gravity. The two are not the same but are very closely related. The above formula holds for any CFT, including for rational CFTs without weakly coupled gravity duals. However, for irrational CFTs at large central charge, it becomes essentially equivalent to the holographic replica wormhole formula in AdS/CFT duality. 

As a first step we note that we can rewrite the integral of the Weil-Peterson symplectic form over the wormhole surface $\Sigma_n$ as follows
\bea
\label{flip}
\int_{\Sigma_n} \Omega_{\mbox{\tiny WP}} \is \frac{k}{4\pi} \int_{\strut_{{}^{\mbox{\scriptsize $\nspc \Sigma_n\! \times\! D_*$}}}} \hspace{-3mm} \!\!\!\! {\rm tr}( \delta A \wedge \delta A)
\, = \, \frac{k}{4\pi} \int_{\strut_{{}^{\mbox{\scriptsize $\nspc D_*\!\! \times\! \Sigma_n$}}}} \hspace{-3mm} \!\!\! {\rm tr}( d {\cal A} \wedge d {\cal A}) 
\label{flip2}
\eea
Here in the first line, $A$ denote a flat gauge potential with $F(A) =0$ along the punctured disk $D_*$ and $\delta$ denotes the exterior derivative along the $\Sigma_n$. In the second line, ${\cal A}$ denote  a flat gauge potential with ${\cal F}({\cal A}) =0$ along $\Sigma_n$, and $d$ denotes the exterior derivative along~$D_*$. To prove the equality, we note that a flat gauge field along $D_*$ can locally be written in the form $A= g^{-1} d g$. Similarly, a flat gauge field along $\Sigma_n$ takes the form ${\cal A} =g^{-1} \delta g$. So $
 \delta A=  [g^{-1} \delta g, g^{-1} d g] \spc = \spc  - d{\cal A}$,
 which establishes the equality \eqref{flip}.
 
We can recast the second integral \eqref{flip2} in a more familiar form by introducing a lagrange multiplier that imposes the flatness constraint ${\cal F}({\cal A}) =0$, and by performing a partial integration on $D_{\nspc \Delta}$. The integral \eqref{flip2} then turns into the action functional of an $SL(2,\mathbb{R})$ Chern-Simons theory defined on the direct product manifold ${\cal M}_3= S^1 \times \Sigma_n$.
\bea
S[{\cal A},A_\theta] \is \frac{k}{4\pi}\oint_{\strut_{{}^{\mbox{\scriptsize $S^1$}}}} \hspace{-2mm}d\theta \int_{\strut_{{}^{\mbox{\scriptsize $\Sigma$}}}}\! {\rm tr}\bigl( {\cal A} \nspc \wedge\nspc \partial_{\smpc \theta}\smpc {\cal A} \, + \, A_{\smpc \theta} \spc {\cal F}({\cal A})\bigr)\\[3mm]
\is \frac{k}{4\pi}\int_{\strut_{{}^{\mbox{\scriptsize $\nspc S^1\! \times\! \Sigma$}}}} \hspace{-5mm} {\rm tr}\bigl( {\bf A} \nspc \wedge\nspc {\bf d A} \, + \, \frac{2}{3} {\bf A} \nspc \wedge\nspc  {\bf A} \nspc \wedge\nspc  {\bf A}\bigr) 
\eea
where ${\bf A} = (A_\theta, {\cal A})$ denotes the three dimensional $SL(2,\mathbb{R})$ gauge field.  Here the integral needs to be performed with the appropriate boundary conditions discussed in the footnote. These are equivalent to the standard asymptotic AdS boundary conditions in AdS/CFT.

To obtain the chiral partition function with fixed conformal weight $\Delta$, we need to impose
\bea
C(A)\nspc \is\nspc {\rm tr}\bigl( {\rm P} \exp \oint_{S^1}\!\nspc{\bf A} \bigr) \spc = \spc C_\Delta
\eea
Note that this restriction is topological: since the three manifold ${\cal M}_3$ is a product manifold, the value of $C(A)$ is constant on $\Sigma_n$. It therefore takes the same value on all boundary components.  The chiral CFT partition function $Z_{\rm \Delta}(\Sigma_n)$ 
can thus be written as a covariant 2+1-D path integral
 \bea
 \label{zchiral}
Z_{\rm \Delta}(\Sigma_n)  \nspc \is \nspc
 \int_{\strut{\raisebox{-3pt}{\tiny $
 {\, \spc C({\bf A})=C_{\Delta}}$}}}
 \hspace{-11mm}\! 
 [d{\bf A}] 
 \; e^{- CS({\bf A})} 
\eea
where the fields and action are defined on $\Sigma_n$ and where we impose appropriate asymptotic boundary conditions on the gauge fields at $\partial \Sigma_n$. In the special case $n\!=\!1$ where $\Sigma_1$ becomes a disk $D$, the chiral partition function \eqref{zchiral} reduces to the standard co-adjoint orbit path integral of the Virasoro group, reproducing the Virasoro character with the corresponding primary weight.  Hence by construction, equations \eqref{zcft}-\eqref{zchiral}  for $\Sigma_n\!=\!D$ give an exact functional integral representation of the thermal CFT partition function. Our proposal is that the same expression \eqref{zcft}-\eqref{zchiral} generalized to higher topology also gives an exact representation of the $n$-fold wormhole partition function $Z_{\rm CFT}(\Sigma_n)$. 

The Einstein action with negative cosmological constant can be written in first order form as the difference of two $SL(2,\mathbb{R})$ Chern-Simons actions, in terms of the linear combinations ${\bf A^a} = e^a +\omega^a$ and $\bar{\bf A}^a = e^a -\omega^a$ of the 3D frame field $e^a$ and spin connection $\omega^a$. Via the above dictionary, we can associate each CS-theory with one chiral half of the dual 2D CFT. This match \cite{HV:1989} between the Einstein action and the geometric action of the Virasoro group is a key element of the AdS${}_3$/CFT${}_2$ correspondence.

\begin{figure}[b]
\begin{center}
\vspace{0cm}
\includegraphics[scale=1.25]{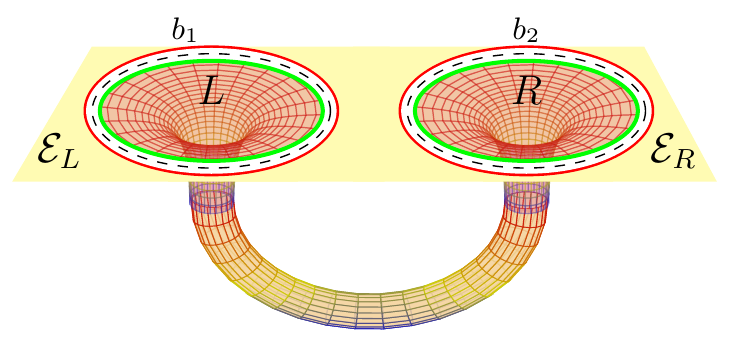}
\end{center}
\vspace{-1mm}
\caption{The two entanglement cuts surrounding the near horizon region of the two-sided black hole introduce edge states on each side of the cut. The combined state of edges  across each cut are described by a boundary state of the holographic CFT.}
\vspace{-2mm}
\end{figure}

\subsection{TMD in 2D CFT}

Now consider two CFTs connected by a wormhole, as shown in figure 1. The wormhole describes the spatial slice of a 3D bulk theory. What is the state that describes the wormhole space-time?  For concreteness, we first consider the case of a 2D CFT that describes the chiral edge modes of a Quantum Hall system. Our construction above then amounts to a derivation of the 3D Chern-Simons action from the phase space path-integral of the edge theory. 
We can think of the wormhole geometry as obtained by cutting space-time along two circles surrounding the left and right regions as shown. 

The cut creates an entanglement boundary between the wormhole region and the environment. 
The physical step of making the entanglement cut amounts to performing a quantum quench that turns off the interaction Hamiltonian that connects the two sides of the cut \cite{Calabrese:2006}\cite{Ryu:2016}. This liberates the edge modes: all edge excitations are no longer gapped out and appear with equal probability in the density matrix of the wormhole space-time region. After tracing out the exterior edge modes associated with the ambient space-time ${\cal E}$, we obtain a reduced CFT density matrix of  the new thermo-mixed double form \cite{Ryu:2016}
\bea
\label{rhocft}
 \rho_{\rm cft}  \, = \, \sum_j \, p_j \, \rho_j^{L} \otimes \rho_j^{R}\qquad \qquad
\rho_j \is \frac{1}{N_{\nspc j}} \sum_{s=1}^{N_{\nspc j}}\,  |j,s\rangle \langle j ,s|
\eea
Here $p_j$ are thermal and $s$ labels the states in the chiral sector labeled by $j$ and the sum runs up to some non-universal level cut-off $N_j = N_{\! j}(\epsilon)$. The R\'enyi entropy von Neumann entropy and mutual information of the above TMD state read
\bea
\tr(\rho_{\rm cft}^n)\! \is \! \sum_j\, \frac{p_j^n}{N_{\nspc j}^{2n-2}\!\!\!\!\!\!\!\!\!} \ \ \  \; ,
\qquad \
S(\rho_{\rm cft}) \smpc = \smpc S_p(\beta)\nspc + \langle  2 \log\nspc N_{\nspc j} \spc \rangle , \quad\  I_{LR}(\rho_{\rm cft}) \spc =\spc S_p(\beta) \, .\quad
\eea
Note that the descendent states are all degenerate. Correspondingly, they are classical and uncorrelated \cite{HV:2021b}. In particular, they do not contribute anything to the entanglement or the mutual information between the {\small $L$} and {\small $R$} sector.

In RCFT,  the level cut-off is naturally set equal to $N_j = N(\epsilon) \, d_j$, with $d_j$ the quantum dimension of the $j$ sector and $N(\epsilon)$ an overall non-universal cut-off scale. This quantum dimension $d_j$ can thus be thought of as counting the multiplicity of the representation $j$.

 In AdS3/CFT2, the gapless modes on the interior edge are identified with the holographic CFT.  In this setting, we do not expect any degeneracies in the spectrum, or that representations occur with high multiplicity. It is then natural to choose the level cut-off to be $j$-independent $N_j = e^{S_0}$. Here ${S_0}$ represents the cut-off scale dependent contribution associated with short-range entanglement across the cut. The universal contribution associated with the local region of the wormhole, on the other hand, should be cut-off independent~\cite{McGough:2013}. Correspondingly, it is natural to decompose the density matrix \eqref{rhocft} as
 \bea
& & \qquad \qquad \qquad \qquad \qquad \ \rho^{\rm old}_{\tMD}  \, = \, \sum_j \, p_j \, |j\rangle_{L}\langle j|  \otimes  |j\rangle_{R}\langle j|\nonumber \\[-3mm]
 \rho_{\rm cft} \is  \rho^{\rm old}_{\tMD} \otimes \rho(\epsilon) \quad {\rm with} \qquad\qquad \qquad  \ \ \\[-3mm]
 & & \qquad \qquad \qquad \qquad \qquad \ \rho(\epsilon)  \, = \, \sum_{s,s'} \, e^{-S_0} |s\rangle_{L}\langle s|  \otimes  |s'\rangle_{R}\langle s'|\nonumber
 \eea
where $\rho^{\rm old}_{\tMD}$ represents the physical mixed state associated with the wormhole geometry.


\begin{thebibliography}{99}
    
\bibitem{replica1} 
  G.~Penington, S.~H.~Shenker, D.~Stanford and Z.~Yang,
{\it Replica wormholes and the black hole interior,}
  arXiv:1911.11977 .
  
\bibitem{replica2} 
  A.~Almheiri, T.~Hartman, J.~Maldacena, E.~Shaghoulian and A.~Tajdini,
{\it Replica Wormholes and the Entropy of Hawking Radiation,}
  arXiv:1911.12333 .

\bibitem{DonHenry} 
  D.~Marolf and H.~Maxfield,
{\it Transcending the ensemble: baby universes, spacetime wormholes, and the order and disorder of black hole information,}
  arXiv:2002.08950. 

\bibitem{Cotler:2016}
J.~S. Cotler, G.~Gur-Ari, M.~Hanada, J.~Polchinski, P.~Saad, S.~H. Shenker,
  D.~Stanford, A.~Streicher, and M.~Tezuka, {\it {Black Holes and Random
  Matrices}},  {\em JHEP} {\bf 05} (2017) 118,
  [\href{http://xxx.lanl.gov/abs/1611.04650}{{\tt 1611.04650}}].

\bibitem{Saad:2018}
P.~Saad, S.~H. Shenker, and D.~Stanford, {\it {A semiclassical ramp in SYK and
  in gravity}},  \href{http://xxx.lanl.gov/abs/1806.06840}{{\tt 1806.06840}}.

\bibitem{Saad:2019}
P.~Saad, S.~H. Shenker, and D.~Stanford, {\it {JT gravity as a matrix
  integral}},  \href{http://xxx.lanl.gov/abs/1903.11115}{{\tt 1903.11115}}.
  
\bibitem{Penington:2019} 
  G.~Penington,
{\it Entanglement Wedge Reconstruction and the Information Paradox,}
  arXiv:1905.08255 .
  
\bibitem{Almheiri:2019} 
  A.~Almheiri, N.~Engelhardt, D.~Marolf and H.~Maxfield,
  {\it The entropy of bulk quantum fields and the entanglement wedge of an evaporating black hole,}
  arXiv:1905.08762 .
  
  \bibitem{Almheiri:2019b} 
  A.~Almheiri, R.~Mahajan, J.~Maldacena and Y.~Zhao,
{\it The Page curve of Hawking radiation from semiclassical geometry,}
  arXiv:1908.10996 .

\bibitem{Witten:1999}
E.~Witten and S.-T. Yau, {\it {Connectedness of the boundary in the AdS/CFT
  correspondence}},  \href{http://xxx.lanl.gov/abs/hep-th/9910245}{{\tt
  hep-th/9910245}}.
\bibitem{maldamaoz}
J.~Maldacena and L.~Maoz, {\it {Wormholes in AdS}},  {\em Journal of High
  Energy Physics} {\bf 2004} (2004), no.~02 053,
  [\href{http://xxx.lanl.gov/abs/hep-th/0401024}{{\tt hep-th/0401024}}].
\bibitem{StanfordWitten:2019}
D.~Stanford and E.~Witten, {\it {JT gravity and the ensembles of random matrix
  theory}},  \href{http://xxx.lanl.gov/abs/1907.03363}{{\tt 1907.03363}}.
\bibitem{Bousso:2020}
R.~Bousso and E.~Wildenhain, {\it Gravity/ensemble duality},  {\em Physical
  Review D} {\bf 102} (2020), no.~6 066005,
  [\href{http://xxx.lanl.gov/abs/2006.16289}{{\tt 2006.16289}}].
\bibitem{Blommaert:2020}
A.~Blommaert,
{\it Dissecting the ensemble in JT gravity,}
[arXiv:2006.13971 [hep-th]].
\bibitem{Cotler2020}
J.~Cotler and K.~Jensen, {\it {AdS$_3 $ gravity and random CFT}},
  \href{http://xxx.lanl.gov/abs/2006.08648}{{\tt 2006.08648}}.
\bibitem{Pollack2020}
J.~Pollack, M.~Rozali, J.~Sully, and D.~Wakeham, {\it Eigenstate thermalization
  and disorder averaging in gravity},  {\em Physical Review Letters} {\bf 125}
  (2020), no.~2 021601, [\href{http://xxx.lanl.gov/abs/2002.02971}{{\tt
  2002.02971}}].
\bibitem{Belin2020}
A.~Belin and J.~de~Boer, {\it {Random statistics of OPE coefficients and
  Euclidean wormholes}},  \href{http://xxx.lanl.gov/abs/2006.05499}{{\tt
  2006.05499}}.
\bibitem{Stanford:2020}
D.~Stanford, {\it {More quantum noise from wormholes}},
  \href{http://xxx.lanl.gov/abs/2008.08570}{{\tt 2008.08570}}.
\bibitem{Kontsevich:1997}
M.~Kontsevich,
{\it Deformation quantization of Poisson manifolds},
Lett. Math. Phys. \textbf{66}, 157-216 (2003),
[arXiv:q-alg/9709040 [math.QA]].
\bibitem{Cattaneo:1999}
A.~S.~Cattaneo and G.~Felder,
{\it A Path integral approach to the Kontsevich quantization formula,}
Commun. Math. Phys. \textbf{212}, 591-611 (2000),
[arXiv:math/9902090 [math]].

\bibitem{Witten:2010}
E.~Witten,
{\it A New Look At The Path Integral Of Quantum Mechanics,}
[arXiv:1009.6032 [hep-th]].

\bibitem{Verlinde:1989}
H.~L.~Verlinde and E.~P.~Verlinde, {\it Conformal field theory and geometric quantization}, in: Proceedings, Superstrings ’89, Trieste, 1989, pp. 422–449.

\bibitem{Blau:1993}
M.~Blau and G.~Thompson,
{\it Derivation of the Verlinde formula from Chern-Simons theory and the G/G model,}
Nucl. Phys. B \textbf{408}, 345-390 (1993),
[arXiv:hep-th/9305010 [hep-th]].

\bibitem{Schaller:1994}
P.~Schaller and T.~Strobl,
{\it Poisson structure induced (topological) field theories,}
Mod. Phys. Lett. A \textbf{9}, 3129-3136 (1994),
[arXiv:hep-th/9405110 [hep-th]].

\bibitem{Hirshfeld:1999}
A.~C.~Hirshfeld and T.~Schwarzweller,
{\it Path integral quantization of the Poisson sigma model,}
Annalen Phys. \textbf{9}, 83-101 (2000),
[arXiv:hep-th/9910178 [hep-th]].

\bibitem{Iliesiu:2019b}
L.~V.~Iliesiu,
{\it On 2D gauge theories in Jackiw-Teitelboim gravity,}
[arXiv:1909.05253 [hep-th]].

\bibitem{HV:2020}
H.~Verlinde,
{\it ER = EPR revisited: On the Entropy of an Einstein-Rosen Bridge,}
[arXiv:2003.13117 [hep-th]].

\bibitem{Takayanagi:2020} 
  A.~Del Campo and T.~Takayanagi,
 {\it Decoherence in Conformal Field Theory,}
  JHEP {\bf 2002}, 170 (2020),
 arXiv:1911.07861.
  

\bibitem{Iliesiu:2019} 
  L.~V.~Iliesiu, S.~S.~Pufu, H.~Verlinde and Y.~Wang,
 {\it An exact quantization of Jackiw-Teitelboim gravity,}
  JHEP {\bf 1911}, 091 (2019)
    
  \bibitem{HV:1989}
H.~L.~Verlinde,
{\it Conformal Field Theory, 2-$D$ Quantum Gravity and Quantization of Teichmuller Space,}
Nucl. Phys. B \textbf{337}, 652-680 (1990)

\bibitem{Cotler:2018}
J.~Cotler and K.~Jensen,
{\it A theory of reparameterizations for AdS$_3$ gravity,}
JHEP \textbf{02}, 079 (2019)
[arXiv:1808.03263 [hep-th]].
  


\bibitem{Kitaev}
A.~Kitaev {\em KITP strings seminar and Entanglement 2015 program
  (\url{http://online.kitp.ucsb.edu/online/entangled15/})}.

\bibitem{MaldacenaStanford:2016}
J.~Maldacena and D.~Stanford, {\it {Remarks on the Sachdev-Ye-Kitaev model}},
  {\em Phys. Rev.} {\bf D94} (2016), no.~10 106002,
  [\href{http://xxx.lanl.gov/abs/1604.07818}{{\tt 1604.07818}}].


    \bibitem{Calabrese:2006}
  P.~Calabrese and J.~L.~Cardy,
{\it Time-dependence of correlation functions following a quantum quench,}
  Phys.\ Rev.\ Lett.\  {\bf 96}, 136801 (2006);
X.L. Qi, H. Katsura, A.W.W. Ludwig,{\it General Relationship between the Entanglement Spectrum and the Edge State Spectrum of Topological Quantum States,}
Phys. Rev. Lett. 108, 196402, (2012)

\bibitem{Ryu:2016} 
  X.~Wen, S.~Matsuura and S.~Ryu,
  {\it Edge theory approach to topological entanglement entropy, mutual information and entanglement negativity in Chern-Simons theories,}
  Phys.\ Rev.\ B {\bf 93}, no. 24, 245140 (2016),
  
\bibitem{McGough:2013}
L. McGough, H. Verlinde, {\it Bekenstein–Hawking entropy as topological entanglement entropy}, J. High Energy Phys.
1311 (2013) 208, arXiv:1308.2342 [hep-th].

\bibitem{HV:2021b} H.~Verlinde, {\it Deconstruncting the Wormhole: Factorization, Entanglement, and Decoherence}, 
[arXiv:2105.0bbbb [hep-th]].
    \end{thebibliography}
\end{document}